\documentclass[aps, prd, reprint, superscriptaddress, amsmath, amssymb, floatfix]{revtex4-2}

\usepackage{graphicx}
\usepackage{epstopdf}
\usepackage{dcolumn}
\usepackage{bm}
\usepackage[utf8]{inputenc}
\usepackage{color,xcolor}
\usepackage{subcaption}
\usepackage{url}
\usepackage{hyperref}
\usepackage{booktabs}
\usepackage{multirow}
\usepackage{tikz}
\usepackage{float}
\usetikzlibrary{arrows.meta,positioning,fit}

\newcommand{\dd}{\mathrm{d}}

\newcommand{\Om}{\Omega_{\rm GW}}
\newcommand{\Taiji}{\mathrm{Taiji}}
\newcommand{\LISA}{\mathrm{LISA}}

\begin{document}

\title{{Effects of Solar Wind Plasma Noise on Stochastic Gravitational Wave Background Searches with a Space Based Interferometer Network}}

\newcommand{\CQUPhys}{Department of Physics, Chongqing University, Chongqing 401331, P.R. China}
\newcommand{\CQUKeyLab}{Chongqing Key Laboratory for Strongly Coupled Physics, Chongqing University, Chongqing 401331, P.R. China}
\newcommand{\CQUInter}{Institute of Advanced Interdisciplinary Studies, Chongqing University, Chongqing 401331, China}
\newcommand{\SUSTechEarth}{Department of Earth and Sciences, Southern University of Science and Technology, Shenzhen 518055, P.R. China}
\newcommand{\NJUAstro}{The School of Astronomy and Space Science at Nanjing University}

\author{Mengfei~Sun}
\thanks{These authors contributed equally to this work.}
\affiliation{\CQUPhys}
\affiliation{\CQUKeyLab}

\author{Borui~Wang}
\thanks{These authors contributed equally to this work.}
\affiliation{\SUSTechEarth}
\affiliation{\NJUAstro}

\author{Jie~Wu}
\affiliation{\CQUPhys}
\affiliation{\CQUKeyLab}

\author{Jin~Li}
\email{cqujinli1983@cqu.edu.cn}
\affiliation{\CQUPhys}
\affiliation{\CQUKeyLab}
\affiliation{\CQUInter}

\author{Shengyi~Ye}
\affiliation{\SUSTechEarth}

\date{July 7, 2026}

\begin{abstract}
The LISA--Taiji dual detector network improves millihertz SGWB sensitivity through cross correlation measurements. Solar wind plasma, however, can generate plasma noise correlated between detectors and bias SGWB cross correlation estimates. We use high time resolution electron density data from Wind/SWE, estimate the solar wind electron density fluctuation spectrum with the Lomb--Scargle method, and propagate the resulting plasma noise to the TDI A/E channels of the LISA--Taiji network. By including finite arm propagation, Taylor frozen flow spatial correlations, and the network overlap reduction response, we compute the SGWB parameter bias induced by interdetector plasma noise. Although the single detector plasma residual is below the reference noise, the component correlated between detectors can enter the SGWB cross correlation estimator directly. Under dual detector scale coverage, the plasma induced parameter bias for a power law SGWB can reach $12.73\%$ of the corresponding Fisher parameter uncertainty. For M2/M3 cosmic string spectra, the bias in $\ln G\mu$ can reach $19.26\%$ of the corresponding Fisher parameter uncertainty for the network configurations, observing times, and frequency bands considered here. These results show that the impact of solar wind plasma noise cannot be assessed from the single detector residual noise level alone. In LISA--Taiji SGWB searches, the interdetector correlated component of this noise can directly affect parameter estimation.
\end{abstract}

\maketitle

\section{Introduction}

Spaceborne laser interferometers extend gravitational wave observations to the millihertz band. The Laser Interferometer Space Antenna (LISA) has a mature mission definition and science program, while Taiji and TianQin provide complementary space based detection concepts \cite{AmaroSeoane2017,LISA2024,HuWu2017,LuoTaiji2020,LuoTianQin2016,MeiTianQin2021}. This band includes massive black hole mergers, extreme mass ratio inspirals, Galactic compact binaries, and several classes of SGWBs. These backgrounds can arise from physical processes in the early Universe or from the superposition of many unresolved astrophysical sources. The cosmic string background is an important cosmological target, and its energy density spectrum carries information about the string tension and loop evolution. Galactic double white dwarfs and the extragalactic compact binary background form the dominant astrophysical foregrounds and enter the cross correlation data together with detector noise \cite{Maggiore2000,Phinney2001,Regimbau2011,CapriniFigueroa2018,ChenHuang2019}. LISA--Taiji SGWB searches must therefore distinguish cosmological signals from astrophysical foregrounds, instrumental noise, and correlated environmental noise.

The effective strain power of an SGWB in a single detector is usually below the instrumental noise, so searches rely on long observations and interdetector cross correlation. Independent detector noises do not form a stable cross spectrum, allowing the common SGWB response to accumulate coherently \cite{AllenRomano1999,RomanoCornish2017,ThraneRomano2013,Christensen2019}. LISA and Taiji have similar millihertz sensitive bands but different heliocentric configurations, so their baselines and overlap reduction functions change the network sensitivity, polarization response, and parameter constraints \cite{WangHan2021,Seto2020,OmiyaSeto2020,CaiNetworks2023}. Existing studies have derived the TDI A/E responses, overlap reduction functions, power law integrated sensitivities, and Fisher estimates for several LISA--Taiji configurations, and have examined power law and cosmic string backgrounds \cite{Wang2023EPJC,Wang2024PRD}. The SGWB sensitivity is therefore jointly determined by detector noise, network geometry, and the cross correlation response. Correlated environmental noise adds a further cross spectrum component and can bias the SGWB observable.

\begin{figure}[tbp]
\centering
\includegraphics[width=\columnwidth]{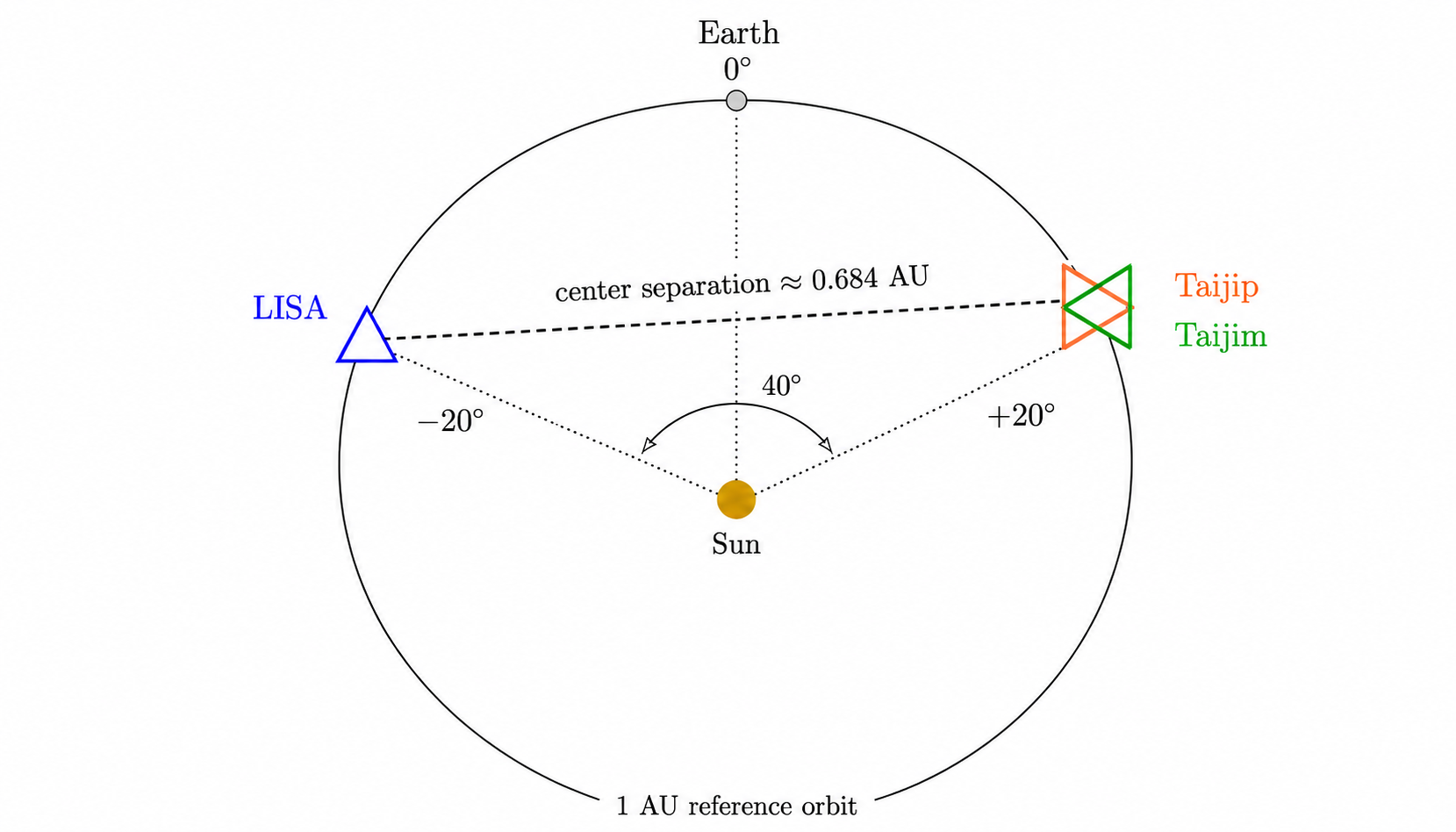}
\caption{Heliocentric geometry of the LISA--Taijip and LISA--Taijim network orientations. The detector centers of LISA and Taiji are located at approximately $-20^\circ$ and $+20^\circ$, respectively, with a center separation of about $0.684$ astronomical unit (AU). The two configurations are drawn in the same coordinates; their main difference is the reversed Taiji triangular orientation.}
\label{fig:geometry}
\end{figure}

{Figure~\ref{fig:geometry} shows the LISA--Taijip and LISA--Taijim configurations. Taijip and Taijim are alternative orientations of one Taiji constellation at the same orbital position, not two simultaneously operating detectors. Both networks have the same center separation, but their different Taiji orientations change the link directions and interdetector correlation response.}

{Ground based SGWB searches face an analogous problem because Schumann resonances excited by global lightning in the Earth ionosphere cavity generate long range correlated magnetic fields. Local magnetic coupling can introduce these fields into the strain channels of widely separated detectors such as LIGO and Virgo \cite{ThraneSchumann2013}. Their impact and mitigation have been studied extensively \cite{CoughlinSchumann2018,MeyersMagnetic2020}. O3 data showed no evidence of significant contamination, although continued monitoring remains necessary as detector sensitivity improves \cite{AbbottO3SGWB2021}.}

Both LISA and Taiji operate in heliocentric orbits, and their interspacecraft laser links pass through the solar wind and interplanetary magnetic field. Free electrons modify the refractive index along each link, so the integrated electron column density fluctuation appears as optical path or arrival time noise in the gravitational wave ranging observable \cite{Smetana2020,Jennrich2021}. For LISA, estimates range from appreciable propagation noise to levels below the dominant noise limits when three dimensional turbulence, Taylor frozen flow, and finite arm averaging are included \cite{Smetana2020,Jennrich2021}. Analyses based on long duration Wind/SWE electron density data and Lomb--Scargle spectral estimation likewise place the Taiji single link and first generation TDI plasma noise below the mission reference noise \cite{Ogilvie1995,OgilvieDesch1997,Xie2024}. TianQin studies have examined related effects in optical path, wave front, pointing, and TDI observables across different orbital configurations \cite{Lu2021,Su2021,Jing2022,Lingfeng2020,Zhou2026}.

{Previous work has also considered several mitigation and monitoring strategies for plasma propagation noise. The BBO mission concept study identified solar wind refractive index fluctuations as correlated noise and proposed an auxiliary radio interferometer to monitor and remove this contribution \cite{PhinneyBBO2004}. Smetana proposed a second laser wavelength to exploit plasma dispersion and separate the plasma contribution from the gravitational wave response \cite{Smetana2020}. For TianQin, first and second generation TDI calculations found residual plasma noise below the secondary noise requirement under normal solar conditions, with moderate suppression below approximately $10^{-2}\,\mathrm{Hz}$ due to correlations among the arm noises \cite{Jing2022}.}

Previous studies mainly compared single link or single detector plasma noise with displacement, pointing, or TDI noise requirements. These results characterize the plasma noise of one detector but do not determine the interdetector cross spectrum used in LISA--Taiji SGWB searches. Even when the residual lies below the noise of each detector, a common solar wind structure can correlate the two sets of links and bias the SGWB cross correlation observable. The magnitude of this contribution depends on the electron density spectrum, spatial correlation scale, finite arm averaging, TDI delay phases, and network geometry.

We investigate the parameter bias caused by interdetector solar wind plasma noise in LISA--Taiji SGWB searches. We infer the electron density fluctuation spectrum from Wind/SWE observations and use finite arm propagation and Taylor frozen flow to construct the plasma noise correlations between links. The LISA--Taiji geometry and TDI A/E responses then map these link correlations into the interdetector channel cross spectra. We quantify the single detector plasma noise and interdetector correlation strength, and evaluate the equivalent SGWB amplitude shift and Fisher parameter bias for power law and M2/M3 cosmic string spectra. {The numerical results in this study are specific to the network configurations formed by LISA and Taiji, while the cross spectral framework can also be applied to other networks of spatially separated space based interferometers.}

Section~II describes the Wind/SWE data and plasma propagation noise model. Section~III introduces the dual detector configuration and TDI noise model. Section~IV presents the SGWB cross correlation response and interdetector plasma noise model. Sections~V and VI present the power law and cosmic string results, respectively, and Section~VII summarizes the conclusions.

\section{Solar Wind Electron Density Data and Plasma Propagation Noise Model}
We first introduce the Wind/SWE electron density data, estimate the electron density power spectrum, and convert it to plasma propagation noise.

\subsection{Wind/SWE electron density data}

\begin{figure}[t]
\centering
\includegraphics[width=\columnwidth]{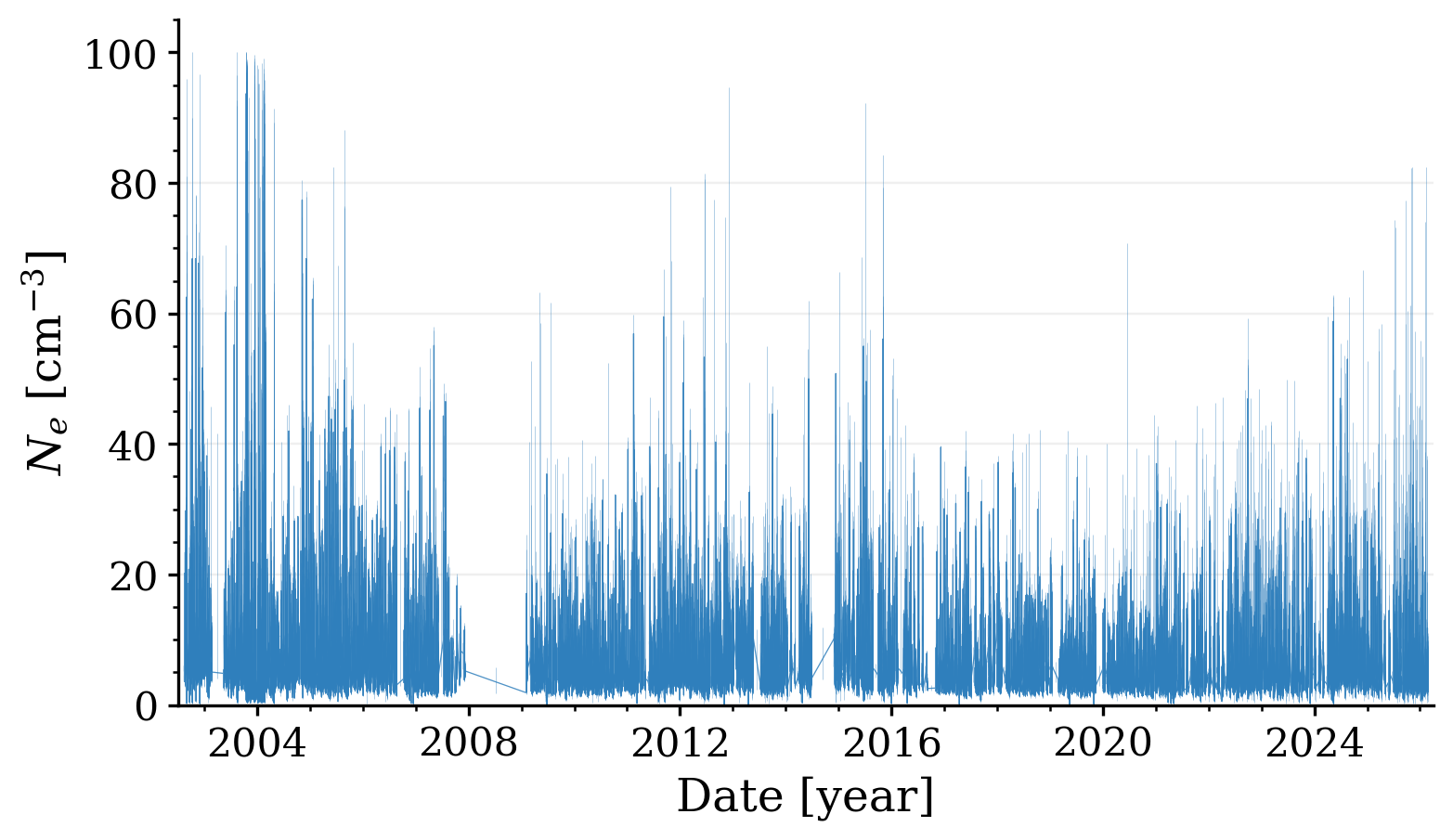}
\caption{Wind/SWE electron density time series. Valid samples from August 2002 to February 2026 are retained, and intermittent sampling is handled directly with Lomb--Scargle spectral estimation.}
\label{fig:wind}
\end{figure}

Wind/SWE has measured the solar wind electron density directly in near Earth space over long intervals \cite{Ogilvie1995,OgilvieDesch1997}. We use the latest observations available when the data were processed, namely high time resolution SWE electron density data from August 16, 2002, to February 26, 2026, retaining $38{,}198{,}016$ valid samples \cite{WindSWEH5Data}. The retained samples correspond to a data coverage fraction of $0.634$ over the selected time span, and their median sampling interval is $12.319\,\mathrm{s}$. Figure~\ref{fig:wind} shows the long term variation of the electron density, and Table~\ref{tab:datafit} summarizes the data range and two band fitting parameters.

The SWE data are not regularly sampled. The dominant sampling intervals range from several seconds to slightly more than ten seconds, but instrument status and telemetry interruptions produce longer gaps. If the data are first interpolated onto a regular grid and then Fourier transformed, the interpolation kernel modifies the high frequency power and long gaps introduce spectral leakage through the window function. We therefore do not fill missing samples. Instead, each observation time $t_j$ and its electron density $N_{e,j}$ are retained. Before power spectrum estimation, we subtract the mean separately for each continuous observing segment and remove anomalous electron density values. Lomb--Scargle periodograms are then computed for each segment on a common frequency grid and combined. This procedure preserves the timing information of uneven sampling and avoids regular grid interpolation effects in the spectral shape \cite{Xie2024}.

\begin{table*}[t]
\centering
\caption{Wind/SWE data and two band fits to the electron density power spectral density (PSD). Here $f_{\rm fill}$ is the data coverage fraction over the time span used in the analysis.}
\label{tab:datafit}
\begingroup
\setlength{\tabcolsep}{10pt}
\begin{tabular}{lccccc}
\hline\hline
Date range & Valid points & $f_{\rm fill}$ & Band/Hz & $\alpha_n$ & $\log_{10}b$\\
\midrule
2002 Aug 16 to 2026 Feb 26 & 38,198,016 & 0.634 & $10^{-6}$ to $10^{-4}$ & $-1.554$ & $-3.018$\\
2002 Aug 16 to 2026 Feb 26 & 38,198,016 & 0.634 & $10^{-4}$ to $10^{-2}$ & $-1.124$ & $-1.507$\\
\hline\hline
\end{tabular}
\endgroup
\end{table*}

\subsection{Electron density power spectrum estimation}

The SWE data contain uneven sampling and long gaps. If a regularly sampled periodogram is used directly, sampling window effects are mixed into the power spectrum estimate. The Lomb--Scargle method does not require prior interpolation of missing samples. Instead, it performs a least squares fit to sine and cosine basis functions at each frequency, making it suitable for the electron density time series used here \cite{Lomb1976,Scargle1982,VanderPlas2018}. For the demeaned sequence $x_j=N_{e,j}-\overline{N}_e$, we use the standard normalized form
\begin{equation}
\begin{aligned}
P_{\rm LS}(\omega)
&=\frac{1}{2s_x^2}\left(P_c+P_s\right),\\
P_c
&=\frac{\left[\sum_jx_j\cos\omega(t_j-\tau)\right]^2}
{\sum_j\cos^2\omega(t_j-\tau)},\\
P_s
&=\frac{\left[\sum_jx_j\sin\omega(t_j-\tau)\right]^2}
{\sum_j\sin^2\omega(t_j-\tau)},\\
\tan(2\omega\tau)
&=\frac{\sum_j\sin(2\omega t_j)}
{\sum_j\cos(2\omega t_j)} .
\end{aligned}
\label{eq:lomb}
\end{equation}
Here $P_c$ and $P_s$ are the cosine and sine projected powers, the time shift $\tau$ orthogonalizes the sine and cosine basis functions at a given frequency, and $s_x^2$ is the sample variance. Equation~\eqref{eq:lomb} defines the Lomb--Scargle least squares spectrum used here \cite{Xie2024}. We then convert the dimensionless $P_{\rm LS}$ into the one sided electron density spectrum $S_{N_e}(f)$, with units of electron density squared per hertz, using the observing duration and the same spectral convention. We use this electron density power spectrum in the subsequent plasma propagation noise calculation.

We model the resulting electron density fluctuation spectrum as a power law,
\begin{equation}
S_{N_e}(f)=b f^{\alpha_n},
\label{eq:nepsd}
\end{equation}
and fit the two bands $10^{-6}$ to $10^{-4}\,\mathrm{Hz}$ and $10^{-4}$ to $10^{-2}\,\mathrm{Hz}$ separately. In logarithmic linear form, Eq.~\eqref{eq:nepsd} becomes
\begin{equation}
\begin{aligned}
y_k&=\log_{10}S_{N_e}(f_k)
:=\alpha_n x_k+c,\\
x_k&=\log_{10}f_k,\qquad c=\log_{10}b .
\end{aligned}
\label{eq:logfit}
\end{equation}
Letting $\bm X_k=(x_k,1)$ and using a frequency bin weight matrix $\bm W$ for the spectral estimation uncertainties at different frequencies, the fitted parameters and covariance are
\begin{equation}
\begin{aligned}
\widehat{\bm\beta}&=(\bm X^{T}\bm W\bm X)^{-1}
\bm X^{T}\bm W\bm y,\\
\operatorname{Cov}(\widehat{\bm\beta})
&=\widehat{s}_{r}^{\,2}(\bm X^{T}\bm W\bm X)^{-1},
\end{aligned}
\label{eq:psdfitcov}
\end{equation}
where $\bm\beta=(\alpha_n,c)$ and $\widehat{s}_{r}^{\,2}$ is the weighted residual variance. This weighted power law fitting procedure determines the spectral indices and normalizations listed in Table~\ref{tab:datafit}.

\begin{figure}[t]
\centering
\includegraphics[width=\columnwidth]{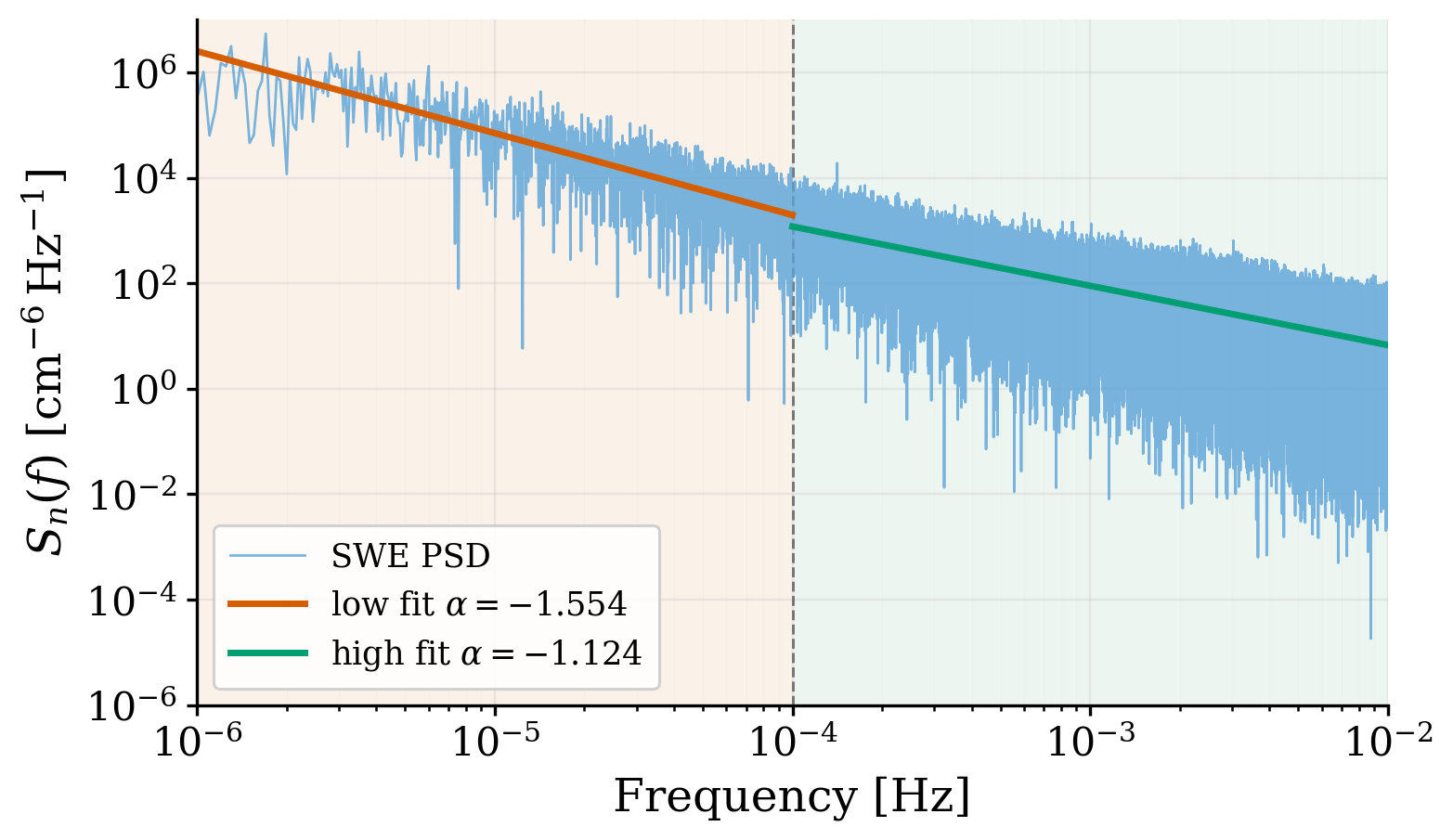}
\caption{Wind/SWE electron density power spectrum and two band power law fits.}
\label{fig:psd}
\end{figure}

To compare the observed spectrum with the commonly used scaling of solar wind turbulence, we write a three dimensional isotropic Kolmogorov model as \cite{Jennrich2021}
\begin{equation}
\left\langle\delta\widetilde n_e^*(\bm k)
\delta\widetilde n_e(\bm k')\right\rangle
:=(2\pi)^3P_0k_0^{11/3}|\bm k|^{-11/3}
\delta^{(3)}(\bm k-\bm k') .
\label{eq:kolmogorov3d}
\end{equation}
Under Taylor's frozen flow assumption $\delta N_e(t)=\delta n_e(-V_{\rm sw}t)$, the wave number and temporal frequency satisfy $k_\parallel=2\pi f/V_{\rm sw}$, yielding the one dimensional scaling
\begin{equation}
S_{N_e}^{\rm K}(f)=
\frac{4\pi^3}{5(2\pi)^{5/3}}P_0k_0^{11/3}
V_{\rm sw}^{2/3}f^{-5/3}.
\label{eq:taylorpsd}
\end{equation}
Here $P_0$ is the amplitude constant of the three dimensional electron density fluctuation spectrum, and $k_0$ is the reference wave number. Equation~\eqref{eq:taylorpsd} provides a theoretical reference for the inertial range of the solar wind. For the calculations below, we use the observed spectrum obtained from the two band fits to the SWE data. This spectrum captures large scale solar wind structures at low frequencies and approaches the scaling expected in the turbulent inertial range at high frequencies. Because the solar wind is multiscale and nonstationary, imposing a single fixed Kolmogorov index would discard part of the observational information \cite{Kolmogorov1941,Taylor1938,Verscharen2019,Abbo2016}. Figure~\ref{fig:psd} shows the full power spectrum and the two band power law fits. We verified the spectral estimate with standard astronomical time series tools \cite{Astropy2013}.

\subsection{Conversion to plasma propagation noise}

When a laser beam passes through solar wind plasma, free electrons reduce the refractive index slightly below its vacuum value. We retain only the cold plasma dispersion term relevant for space laser ranging. Neglecting collisions and magnetization corrections, and taking the laser angular frequency to satisfy $\omega\gg\omega_p$, the Appleton--Hartree relation reduces to the cold plasma refractive index \cite{Hutchinson2002,Smetana2020,Xie2024}
\begin{equation}
n^2=1-\frac{\omega_p^2}{\omega^2},\qquad
\omega_p^2=\frac{N_e e^2}{\epsilon_0m_e}.
\label{eq:refractive}
\end{equation}
For solar wind electron densities and laser frequencies, $\omega_p^2/\omega^2\ll1$. The first order expansion of the refractive index is
\begin{equation}
n\simeq1-\frac{\omega_p^2}{2\omega^2}
:=1-\frac{e^2N_e}{2\epsilon_0m_e\omega^2}.
\label{eq:refractive_expansion}
\end{equation}
Refractive index fluctuations accumulate along the laser link and form optical path fluctuations. For a link of length $L$, the equivalent optical path induced by electron density fluctuations is
\begin{equation}
\begin{aligned}
\delta L_p(t)
&=-\chi(\omega)\int_0^L\delta N_e[\bm r(s),t]\,\dd s,\\
\chi(\omega)
&=\frac{e^2}{2\epsilon_0m_e\omega^2}.
\end{aligned}
\label{eq:path}
\end{equation}
Equation~\eqref{eq:path} shows that laser ranging responds to the electron column density fluctuation integrated over the whole link, rather than to the local density near a single point close to the detector. If the density fluctuation is completely in phase over the full arm, one obtains \cite{Smetana2020}
\begin{equation}
S_L^{\rm coh}(f)=(L\chi)^2S_{N_e}(f).
\label{eq:smetana_transfer}
\end{equation}
The fully coherent arm approximation provides an intuitive scaling relation, but it does not include phase averaging along the finite arm. When solar wind structures sweep across the link with the background flow, different positions along the link correspond to different frozen flow phases, reducing the net contribution of electron density fluctuations to the single link optical path. Starting from the three dimensional Kolmogorov spectrum and frozen flow geometry, the single arm transfer function can be approximated as \cite{Jennrich2021}
\begin{equation}
\begin{aligned}
S_L^{\rm arm}(f)
&\simeq
(L\chi)^2\frac{25}{9}\beta^{5/3}
\left(\frac{V_{\rm sw}}{2\pi Lf}\right)S_{N_e}(f),\\
\frac{\sqrt{3}}{2}
&\lesssim\beta\lesssim1 .
\end{aligned}
\label{eq:jennrich_transfer}
\end{equation}
Here $\beta$ describes the geometric orientation of the link relative to the solar wind. The factor $V_{\rm sw}/(2\pi Lf)$ comes from finite arm averaging and further suppresses high frequency fluctuations. This frequency dependent suppression is the key difference between Eq.~\eqref{eq:jennrich_transfer} and the fully coherent arm approximation. We use Eq.~\eqref{eq:jennrich_transfer} to estimate the single link noise level. The interdetector plasma noise is computed in Sec.~IV with a double path integral, without approximating the full arm as a completely coherent sampling point.

\begin{figure}[t]
\centering
\includegraphics[width=\columnwidth]{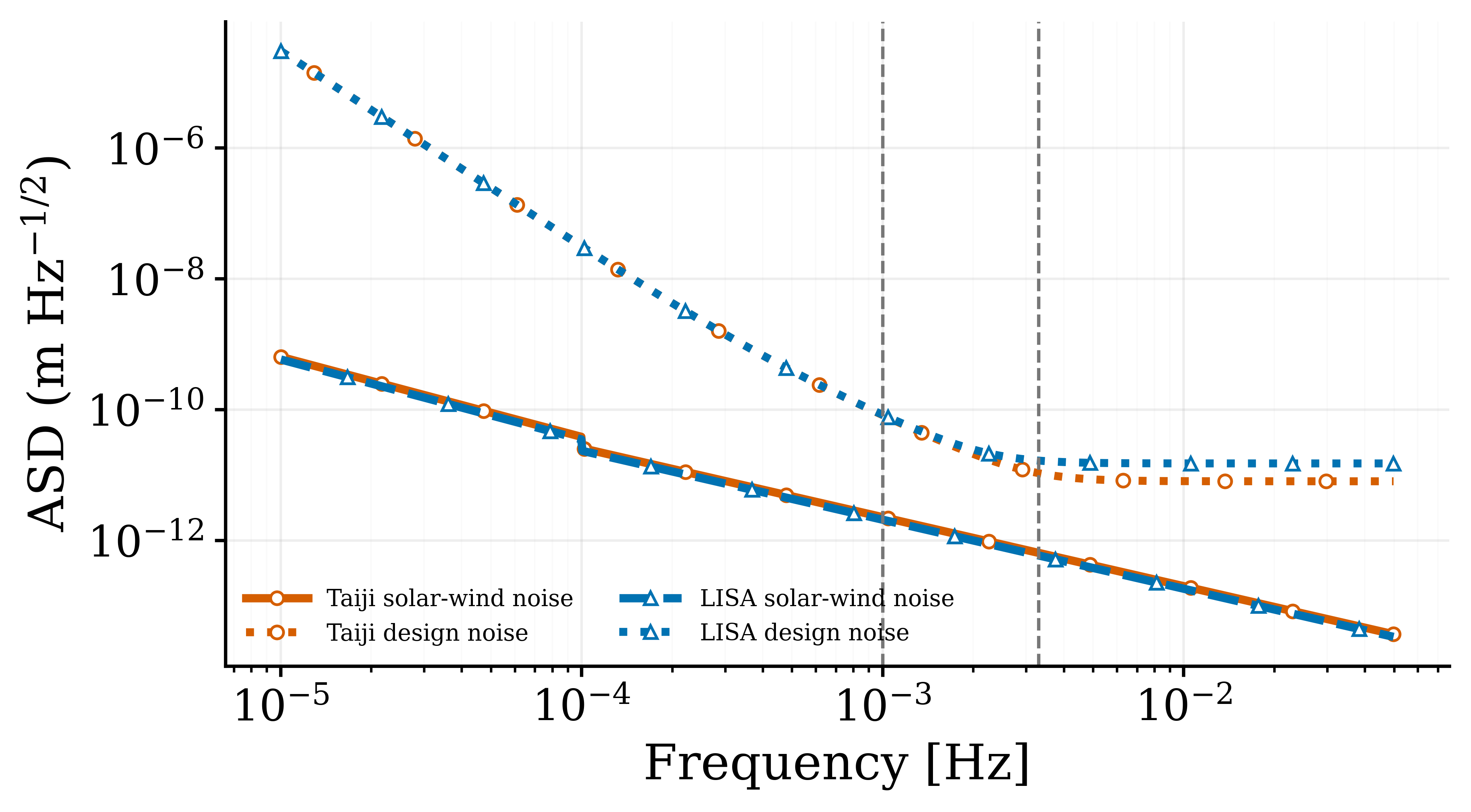}
\caption{Single link plasma displacement noise and reference noise for Taiji and LISA. The vertical lines mark $1\,\mathrm{mHz}$ and $3.3\,\mathrm{mHz}$. Circles and upward-pointing triangles identify Taiji and LISA, respectively. The plasma-noise curves are solid for Taiji and dashed for LISA; dotted curves show the reference noises.}
\label{fig:linknoise}
\end{figure}

Refractive index fluctuations can also change the laser propagation direction through transverse gradients. The ray equation in geometric optics is \cite{Zhou2026}
\begin{equation}
\frac{\dd}{\dd s}\left(n\frac{\dd\bm r}{\dd s}\right)
=\bm\nabla n .
\label{eq:raytrace}
\end{equation}
Equation~\eqref{eq:raytrace} can be used to calculate plasma induced transverse pointing deflection \cite{Zhou2026}. Here we focus on the longitudinal column density optical path fluctuation described by Eq.~\eqref{eq:path} and do not include the transverse pointing term as an additional displacement noise. The two effects arise from the same refractive index perturbation, but they affect optical path and pointing measurements separately. Figure~\ref{fig:linknoise} compares the optical path noise used here with the Taiji and LISA single link reference noises.

\section{Dual Detector Configuration and TDI Noise Model}

We next specify the LISA--Taiji geometry, A/E channel response, instrumental noise model, and propagation of plasma link noise into TDI channels.

\subsection{Detector configuration and noise model}

Taiji and LISA both use approximately equilateral triangular laser interferometer configurations, with arm lengths of about $3.0\times10^6\,\mathrm{km}$ and $2.5\times10^6\,\mathrm{km}$, respectively. We consider the LISA--Taijip and LISA--Taijim network orientations \cite{Wang2023EPJC}. Figure~\ref{fig:geometry} shows the dual detector geometry and intercenter baseline of these two configurations. The orbital orientation of the network determines the frequency structure of the ORF and the geometric response of the two sets of links to the same solar wind structure \cite{WangHan2021,Wang2023EPJC,Wang2024PRD}.

For a detector with arm length $L_d$, the characteristic transfer frequency is
\begin{equation}
f_{*,d}=\frac{c}{2\pi L_d}.
\label{eq:fstar}
\end{equation}
When $f\ll f_{*,d}$, the detector can be approximated as two orthogonal interferometric channels in the long wavelength limit. When the frequency approaches or exceeds $f_{*,d}$, the finite light travel time produces oscillations and nulls in the response. The approximate A/E channel response is \cite{Wang2023EPJC}
\begin{align}
\mathcal R_{A,E}(f)&\simeq
\frac{9}{20}|W(f)|^2
\left[1+\left(\frac{3f}{4f_*}\right)^2\right]^{-1},\\
W(f)&=1-\exp(-2if/f_*).
\label{eq:ae_response}
\end{align}
Our numerical response retains the actual Taiji and LISA arm lengths and network geometry. Equation~\eqref{eq:ae_response} is used only to illustrate the low frequency normalization of the A/E channels and the gradual weakening of the response at high frequencies.

Instrumental noise consists of acceleration noise and optical metrology noise. For detector $d\in\{\Taiji,\LISA\}$, let the A/E channel instrumental power spectrum be $P_{I,d}^{\rm inst}(f)$. We use the same total noise power spectra in the network SGWB response and in all cross spectrum weights,
\begin{equation}
P_{I,d}(f)=P_{I,d}^{\rm inst}(f)+P_{I,d}^{\rm plasma}(f).
\end{equation}
For equal arm first generation TDI, the analytic A/E channel noise has the form \cite{Wang2023EPJC}
\begin{equation}
N_{A,E}(f)=|W(f)|^2\left[6N_o(f)+24N_a(f)\right],
\label{eq:ae_noise}
\end{equation}
where $N_o$ and $N_a$ denote the optical metrology noise and acceleration noise normalized to the link measurements \cite{Wang2023EPJC,Prince2002}. The two fundamental noises enter the A/E channels through the same delay polynomials but with different numerical coefficients. We adopt the acceleration noise and optical metrology noise parameters from the literature and do not refit $N_o$ or $N_a$. When electron density structures mainly affect the links of a single detector, the plasma term is much smaller than the instrumental term. This notation gives a common noise normalization for the single detector noise level, cross correlation likelihood, and network Fisher weights.

\subsection{TDI channels and single detector noise levels}

Time delay interferometry suppresses laser frequency noise through delayed combinations \cite{TintoEstabrook2002,Prince2002,Vallisneri2005,TintoHartwig2018,TintoDhurandhar2021}. The orthogonal channels are defined from Michelson type $X,Y,Z$ combinations as
\begin{equation}
A=\frac{Z-X}{\sqrt2},\quad
E=\frac{X-2Y+Z}{\sqrt6},\quad
T=\frac{X+Y+Z}{\sqrt3}.
\label{eq:aet}
\end{equation}
The delay operator $\mathcal D_{ij}$ acting on a link observable $x(t)$ is defined by
\begin{equation}
\mathcal D_{ij}x(t)=x[t-L_{ij}/c],
\qquad
\widetilde{\mathcal D}_{ij}(f)=
\exp[-2\pi ifL_{ij}/c].
\label{eq:delayop}
\end{equation}
Any first generation TDI combination can be written as a delay polynomial linear combination of the link measurement vector $\bm y_d$,
\begin{equation}
\widetilde q_{I,d}(f)=
\bm D_{I,d}^{\,T}(f)\widetilde{\bm y}_d(f),
\qquad I\in\{X,Y,Z,A,E,T\}.
\label{eq:tdilinear}
\end{equation}
Existing Taiji single detector studies used the TDI $\alpha$ and $X$ combinations as examples to expand the delay ordering of each link and obtain the residual plasma PSD \cite{Xie2024}. Equation~\eqref{eq:tdilinear} expresses the same relation in matrix notation. Each component of $\bm D_{I,d}$ is a sum or difference of several $\widetilde{\mathcal D}_{ij}$ terms and therefore retains the propagation direction and light travel time phase. If the link spectral matrix is denoted by $\bm S_d^{\rm link}$, then the single detector TDI auto spectrum is
\begin{equation}
S_{I,d}(f)=
\bm D_{I,d}^{\,T}\bm S_d^{\rm link}
\bm D_{I,d}^{\,*}.
\label{eq:tdiauto}
\end{equation}

\begin{figure}[t]
\centering
\includegraphics[width=\columnwidth]{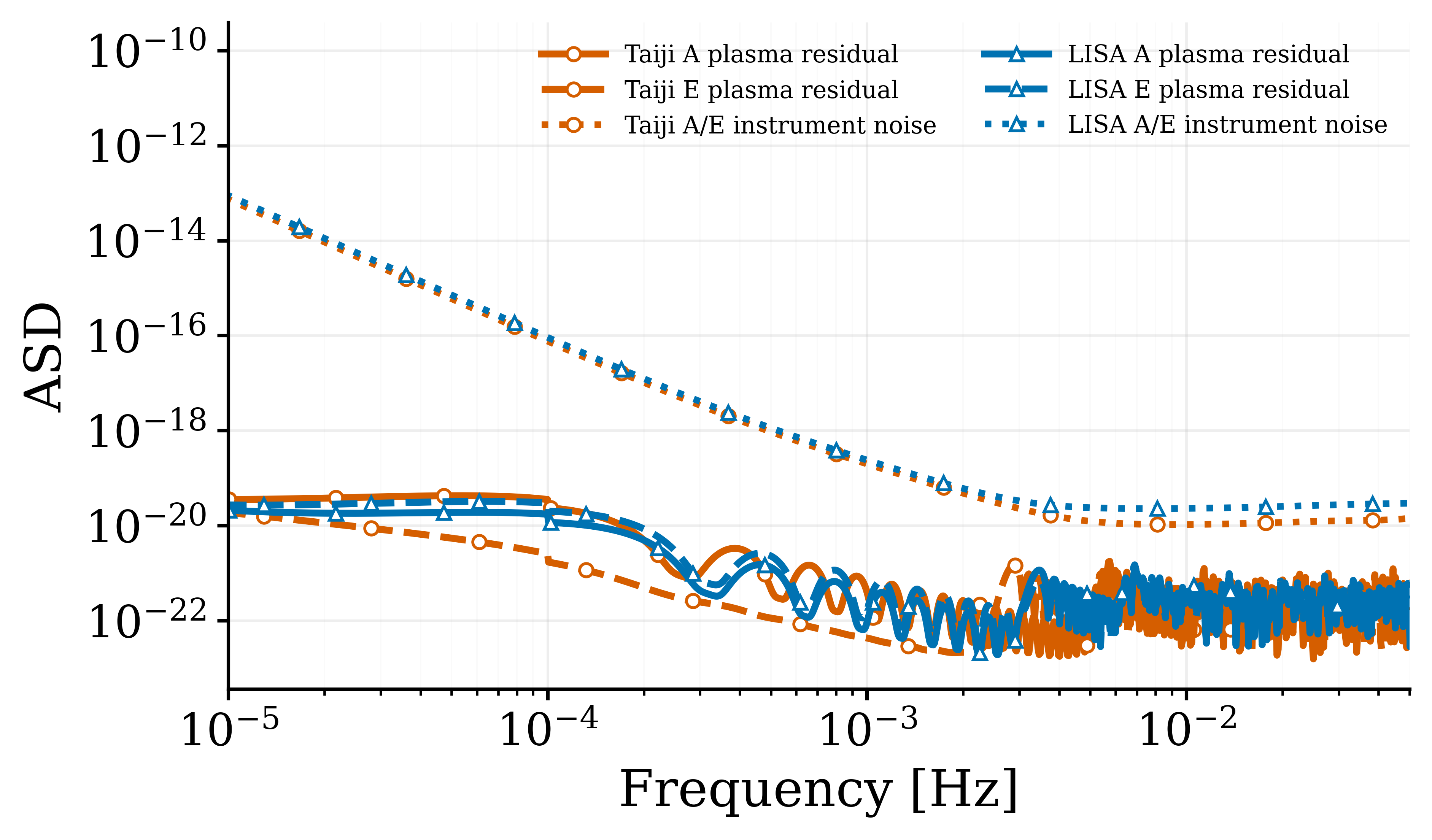}
\caption{Single detector A/E channel plasma residuals and instrumental equivalent strain noises for Taiji and LISA. Circles and upward-pointing triangles identify Taiji and LISA, respectively. Solid and dashed curves show the A- and E-channel plasma residuals, respectively, and dotted curves show the instrumental noise.}
\label{fig:tdi}
\end{figure}

We focus on the A/E channels, where the millihertz SGWB response is relatively strong, and retain the propagation phase of each plasma link before forming the TDI combinations. Finite arm integration describes phase averaging along a link, while the first generation TDI delay polynomial maps the link noise into the A/E channels \cite{Otto2012,Xie2024,Jing2022}. Figure~\ref{fig:tdi} and Table~\ref{tab:budget} summarize the single detector residual noise.

\begin{table*}[t]
\centering
\caption{Selected frequency results for the single detector plasma noise level. Here $A_{\rm pl}$ and $A_{\rm inst}$ denote the plasma amplitude spectral density (ASD) and instrumental ASD in the corresponding detector and channel, respectively, and the last column reports their ratio. The frequency $1\,\mathrm{mHz}$ is a representative millihertz point, and $3.3\,\mathrm{mHz}$ is where the ratio between the single link plasma noise and reference noise is relatively large. The single link rows correspond to displacement ASD, while the A/E rows correspond to response corrected equivalent strain ASD.}
\label{tab:budget}
\begingroup
\setlength{\tabcolsep}{10pt}
\begin{tabular}{llcccc}
\hline\hline
Detector & Channel & $f/{\rm Hz}$ & $A_{\rm pl}$ & $A_{\rm inst}$ & $A_{\rm pl}/A_{\rm inst}$\\
\midrule
Taiji & single link & $1.0\times10^{-3}$ & $1.40\times10^{-12}$ & $8.22\times10^{-11}$ & $1.70\times10^{-2}$\\
Taiji & single link & $3.3\times10^{-3}$ & $3.59\times10^{-13}$ & $1.06\times10^{-11}$ & $3.37\times10^{-2}$\\
LISA & single link & $3.3\times10^{-3}$ & $3.27\times10^{-13}$ & $1.66\times10^{-11}$ & $1.98\times10^{-2}$\\
Taiji & A & $3.3\times10^{-3}$ & $4.23\times10^{-22}$ & $1.98\times10^{-20}$ & $2.14\times10^{-2}$\\
Taiji & E & $3.3\times10^{-3}$ & $2.56\times10^{-22}$ & $1.98\times10^{-20}$ & $1.29\times10^{-2}$\\
LISA & E & $3.3\times10^{-3}$ & $2.81\times10^{-22}$ & $3.02\times10^{-20}$ & $9.31\times10^{-3}$\\
\hline\hline
\end{tabular}
\endgroup
\end{table*}

\section{SGWB Cross Correlation Response and Interdetector Plasma Noise Model}

We now construct the SGWB cross correlation response, finite arm link integration, frozen flow spatial correlation function, and TDI interdetector spectra.

\subsection{SGWB cross correlation response}

The gravitational wave is written as a superposition of plane waves,
\begin{multline}
h_{ab}(t,\bm x)=
\sum_{P=+,\times}\int_{-\infty}^{\infty}\dd f
\int_{S^2}\dd\widehat{\bm n}\,
\widetilde h_P(f,\widehat{\bm n})e_{ab}^{P}(\widehat{\bm n})\\
\times
\exp\left[2\pi if\left(t-\frac{\widehat{\bm n}\cdot\bm x}{c}\right)\right].
\label{eq:gwplane}
\end{multline}
For a stationary, isotropic, unpolarized, and Gaussian SGWB, the Fourier amplitudes satisfy
\begin{align}
&\left\langle\widetilde h_P(f,\widehat{\bm n})
\widetilde h_{P'}^*(f',\widehat{\bm n}')\right\rangle \notag\\
&\quad=\frac{1}{16\pi}\delta_{PP'}\delta(f-f')
\delta^{(2)}(\widehat{\bm n},\widehat{\bm n}')S_h(f).
\label{eq:sgwbstat}
\end{align}
We follow the standard power law energy density parameterization used in SGWB searches and describe an isotropic power law background by its amplitude at the reference frequency $f_{\rm ref}=1\,\mathrm{mHz}$ and a spectral index \cite{Mandic2012,RomanoCornish2017,AbbottO3SGWB2021},
\begin{equation}
\Om(f)=\Omega_{\rm ref}\left(\frac{f}{f_{\rm ref}}\right)^\alpha,
\qquad
S_h(f)=K(f)\Om(f),
\end{equation}
where
\begin{equation}
K(f)=\frac{3H_0^2}{4\pi^2 f^3}.
\label{eq:kf}
\end{equation}
We use this A/E channel SGWB power spectrum normalization \cite{Wang2024PRD,SmithCaldwell2019}. For Taiji channel $I$ and LISA channel $J$, the ORF is defined as
\begin{multline}
\gamma_{IJ}(f)=\frac{5}{8\pi}
\sum_P\int_{S^2}\dd\widehat{\bm n}\,
F_{I,\Taiji}^{P}(f,\widehat{\bm n})
F_{J,\LISA}^{P*}(f,\widehat{\bm n})\\
\times
\exp\left(\frac{2\pi if}{c}
\widehat{\bm n}\cdot\Delta\bm x\right),
\label{eq:orf}
\end{multline}
where $\Delta\bm x=\bm x_{\Taiji}-\bm x_{\LISA}$, and $F_{I,d}^{P}$ includes the detector tensor, finite arm response, and TDI delay response. We adopt the same normalization convention for the A/E channel responses and ORFs as in Refs.~\cite{Wang2023EPJC,Wang2024PRD}. Under this convention, the SGWB cross spectrum in the network channels is given by Eq.~\eqref{eq:crossgw}, where $K(f)$ is defined by Eq.~\eqref{eq:kf} and $\gamma_I(f)$ is the LISA--Taiji ORF for the corresponding A/E channel. In mirror symmetric geometries, the ORFs of the $AE$ and $EA$ channels are suppressed in ideal configurations; the main text therefore uses the corresponding $AA$ and $EE$ cross correlations.

For a network channel $I\in\{A,E\}$, the observed cross spectrum is
\begin{equation}
C_I^{\rm obs}=C_I^{\rm SGWB}+C_I^{\rm plasma}+C_I^{\rm noise},
\end{equation}
\begin{equation}
C_I^{\rm SGWB}(f)=K(f)\gamma_I(f)\Om(f),
\label{eq:crossgw}
\end{equation}
where $\gamma_I(f)$ is the ORF of the corresponding LISA--Taiji channel. Let
\[
C_I(f)=C_I^{\rm SGWB}(f)+C_I^{\rm plasma}(f).
\]
The frequency domain covariance matrix for the two detectors in one A/E channel is
\begin{equation}
\bm{\mathcal C}_I(f)=
\begin{pmatrix}
P_{I,\Taiji}(f) &
C_I(f)\\
C_I^*(f) &
P_{I,\LISA}(f)
\end{pmatrix}.
\label{eq:channelcov}
\end{equation}
For ideally independent instrumental noises, the expectation value of the cross term is zero. The plasma term is specified by the spatial correlation model below. In the weak signal limit $|C_I|^2\ll P_{I,\Taiji}P_{I,\LISA}$, the variance of the cross spectrum estimator is mainly controlled by the product of the two auto spectra. Section~V uses this limit for the cross correlation Fisher integral. Because ORF zeros may cause the pointwise ratio $C_I^{\rm plasma}/C_I^{\rm SGWB}$ to diverge, we use the frequency integrated $\Delta\Om$ and Fisher parameter bias as diagnostics rather than a pointwise ratio.

\subsection{Interdetector plasma noise model}

We extend the single link plasma propagation noise to the LISA--Taiji dual detector link matrix and obtain the complex cross spectra in the A/E channels after TDI combination.

\paragraph*{Finite arm integration model.}
For links $a$ and $b$, the plasma optical path cross spectrum is a double integral over the two propagation paths,
\begin{equation}
S_{ab}^{\rm plasma}(f)=\chi^2 S_{N_e}(f)
\int_0^{L_a}\dd s\int_0^{L_b}\dd s'\,
\Gamma_n[\Delta\bm r(s,s'),f].
\label{eq:double}
\end{equation}
The link path is parameterized as
\begin{equation}
\bm r_a(s)=\bm x_{a,0}+s\widehat{\bm u}_a,\qquad
0\le s\le L_a ,
\label{eq:linkpath}
\end{equation}
so $\Delta\bm r(s,s')=\bm r_a(s)-\bm r_b(s')$. Here $\Gamma_n$ is the normalized electron density correlation kernel. The auto spectrum follows from $a=b$, while the cross spectrum retains both amplitude decay and propagation phase. The auto and cross spectra for all links form a Hermitian matrix,
\begin{equation}
\left[\bm S^{\rm link}(f)\right]_{ab}
:=S_{ab}^{\rm plasma}(f),\qquad
\bm S^{\rm link}=\bm S^{\rm link\dagger}.
\label{eq:linkmatrix}
\end{equation}
A physical correlation matrix must be positive semidefinite, so every link pair satisfies
\begin{equation}
|S_{ab}^{\rm plasma}(f)|^2
\le S_{aa}^{\rm plasma}(f)S_{bb}^{\rm plasma}(f).
\label{eq:cauchylink}
\end{equation}
Equation~\eqref{eq:cauchylink} ensures that the normalized correlation strength at the link level is not larger than unity. Equation~\eqref{eq:double} directly includes the arm lengths, arm directions, and contributions from different position pairs along the links, extending the single link responses in Eqs.~\eqref{eq:smetana_transfer} and \eqref{eq:jennrich_transfer} to the double link and interdetector case.

\paragraph*{Spatial correlation function and frozen flow model.}
The solar wind transports turbulent structures outward with mean speed $V_{\rm sw}$. Taylor's frozen flow approximation maps temporal frequency to the spatial scale $\lambda_{\rm sw}$ \cite{Taylor1938,Jennrich2021}, where
\[
\lambda_{\rm sw}=V_{\rm sw}/f .
\]
We use the anisotropic correlation kernel
\begin{equation}
\Gamma_n=\exp\left[-\frac{|\Delta r_B|}{L_\parallel}
-\frac{\Delta r_\perp^2}{2L_\perp^2}\right]
\exp\left[-\frac{2\pi i f\Delta r_{\rm sw}}{V_{\rm sw}}\right].
\label{eq:kernel}
\end{equation}
Here $\widehat{\bm b}$ denotes the anisotropic correlation axis set by the Parker spiral, $\widehat{\bm v}_{\rm sw}$ denotes the radial advection direction of the solar wind, and
\begin{equation}
\begin{aligned}
\Delta r_B&=\Delta\bm r\cdot\widehat{\bm b},\\
\Delta r_{\rm sw}&=\Delta\bm r\cdot\widehat{\bm v}_{\rm sw},\\
\Delta r_\perp^2&=|\Delta\bm r|^2-\Delta r_B^2.
\end{aligned}
\label{eq:separation}
\end{equation}
The quantities $L_\parallel$ and $L_\perp$ control the correlation lengths along the Parker spiral correlation axis and in the perpendicular direction, respectively. The second exponential in Eq.~\eqref{eq:kernel} retains the complex phase generated when radial solar wind advection maps a spatial displacement to a time delay. At fixed $f$, the correlation amplitude decays with the separation between two points. At fixed separation, the complex phase rotates with frequency. These two effects jointly determine the envelope and frequency structure of the cross spectrum.

\begin{figure}[t]
\centering
\includegraphics[width=\columnwidth]{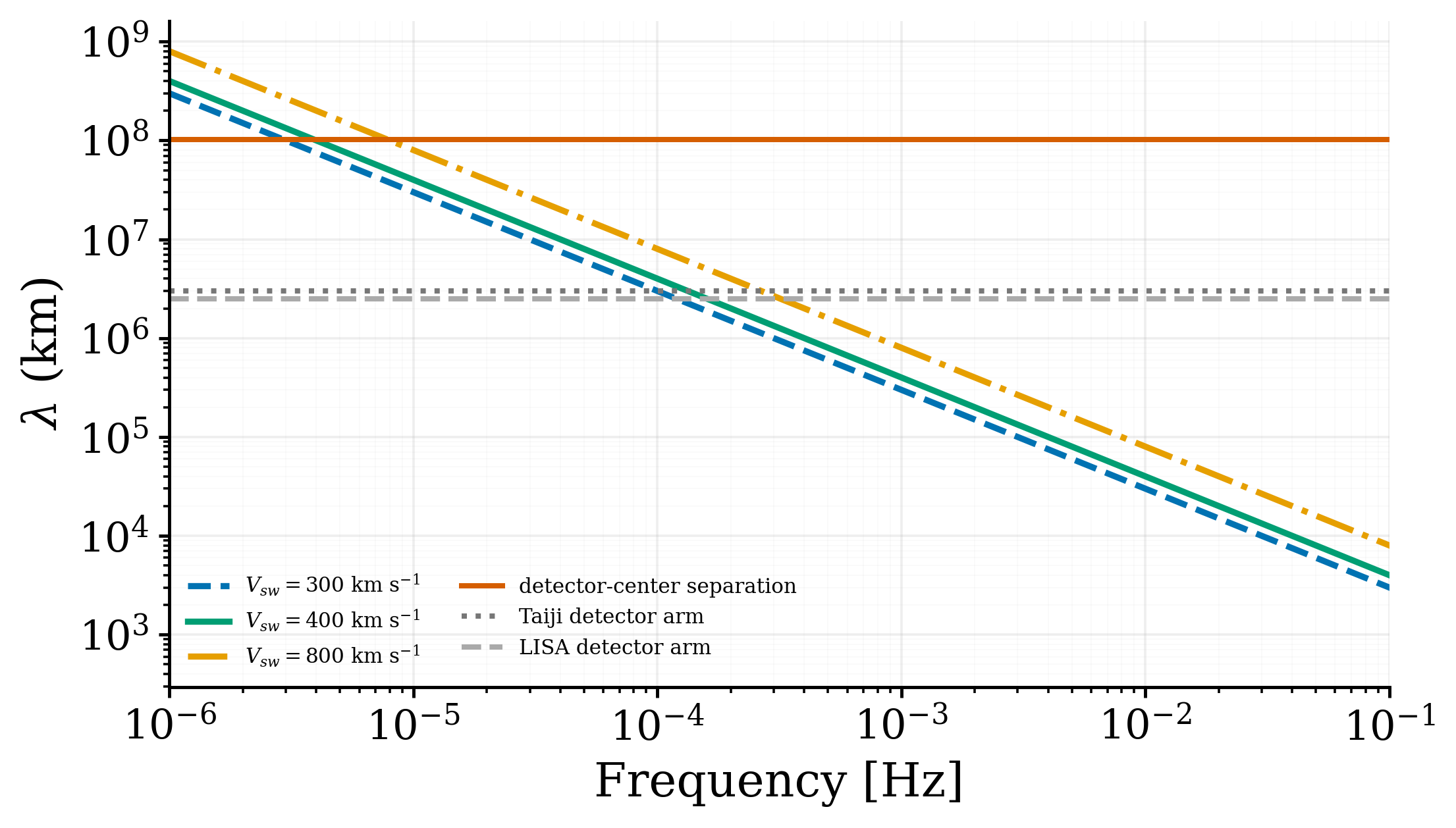}
\caption{Comparison between frozen flow spatial scales and the LISA--Taiji dual detector geometric scales. Millihertz structure scales are much smaller than the detector center separation, so interdetector correlations are strongly suppressed in SDC.}
\label{fig:scale}
\end{figure}

We consider two correlation scale regimes to characterize the spatial coverage of solar wind electron density structures over the LISA--Taiji links. The first is single detector scale coverage (SDC), with $L_\parallel=10^8\,\mathrm{km}$, $L_\perp=10^7\,\mathrm{km}$, and $V_{\rm sw}=400\,\mathrm{km\,s^{-1}}$. This case describes plasma structures that can affect the links of a single detector but are unlikely to cover both Taiji and LISA, which are separated by about $0.684\,\mathrm{AU}$. The second is dual detector scale coverage (DDC), with $L_\parallel=10^9\,\mathrm{km}$ and $L_\perp=10^8\,\mathrm{km}$. {Multipoint spacecraft observations have identified the same magnetic cloud and corotating interaction region complex at spacecraft separated by about $0.7\,\mathrm{AU}$, corresponding to approximately $40^\circ$ \cite{Farrugia2011}. Correlations in electron core density have also been reported over longitudinal separations of about $30^\circ$ after accounting for the solar wind propagation time \cite{Opitz2010}. These observations indicate that large solar wind structures may extend across widely separated heliospheric regions. We therefore define DDC as a large scale correlation regime that covers both LISA and Taiji and use it to assess the resulting interdetector plasma noise. The DDC correlation lengths are model parameters for this regime and are not inferred from the single point Wind/SWE observations.} The Parker spiral angle determines the geometric relation between the correlation axis $\widehat{\bm b}$ and the detector baseline and link directions, while the frozen flow phase is set by the radial solar wind velocity direction $\widehat{\bm v}_{\rm sw}$. Figure~\ref{fig:scale} compares the frozen flow scale, detector arm lengths, and intercenter separation. At $1\,\mathrm{mHz}$, for example, $\lambda_{\rm sw}\simeq4\times10^5\,\mathrm{km}$, which is smaller than a single detector arm length and much smaller than the LISA--Taiji intercenter distance. Interdetector plasma correlations are therefore strongly suppressed by spatial decoherence in SDC.

\paragraph*{TDI interdetector spectra.}
Let $D_{I,d}^a(f)$ be the frequency domain coefficient of detector $d$ and TDI channel $I$ for link vector component $a$. The interdetector plasma spectrum is then
\begin{equation}
\begin{aligned}
C_I^{\rm plasma}(f)
&=\sum_{a\in\Taiji}\sum_{b\in\LISA}
D_{I,\Taiji}^a(f)D_{I,\LISA}^{b*}(f)\\
&\quad \times S_{ab}^{\rm plasma}(f).
\end{aligned}
\label{eq:tdicross}
\end{equation}
Using a block matrix of link spectra, Eq.~\eqref{eq:tdicross} can be written more compactly as
\begin{equation}
C_{IJ}^{\rm plasma}(f)=
\bm D_{I,\Taiji}^{\,T}
\bm S_{\Taiji\LISA}^{\rm link}
\bm D_{J,\LISA}^{\,*},
\label{eq:tdicrossmatrix}
\end{equation}
and the TDI auto spectra of the two detectors are
\begin{equation}
S_{I,d}^{\rm plasma}(f)=
\bm D_{I,d}^{\,T}\bm S_{dd}^{\rm link}
\bm D_{I,d}^{\,*}.
\label{eq:tdiautoplasma}
\end{equation}
The corresponding coherence, or normalized correlation strength, is defined by
\begin{equation}
\rho_I^{\rm plasma}(f)=
\frac{|C_I^{\rm plasma}(f)|}
{\sqrt{S_{I,\Taiji}^{\rm plasma}(f)S_{I,\LISA}^{\rm plasma}(f)}}.
\label{eq:coherence}
\end{equation}
Positive semidefiniteness of the spectral matrix implies $0\le\rho_I^{\rm plasma}\le1$. This quantity measures only the normalized linear correlation strength of the plasma components in the two TDI data streams. It does not include the SGWB ORF, instrumental noise weighting, or parameter response, and therefore differs physically from the SGWB parameter bias.

\begin{figure}[t]
\centering
\includegraphics[width=\columnwidth]{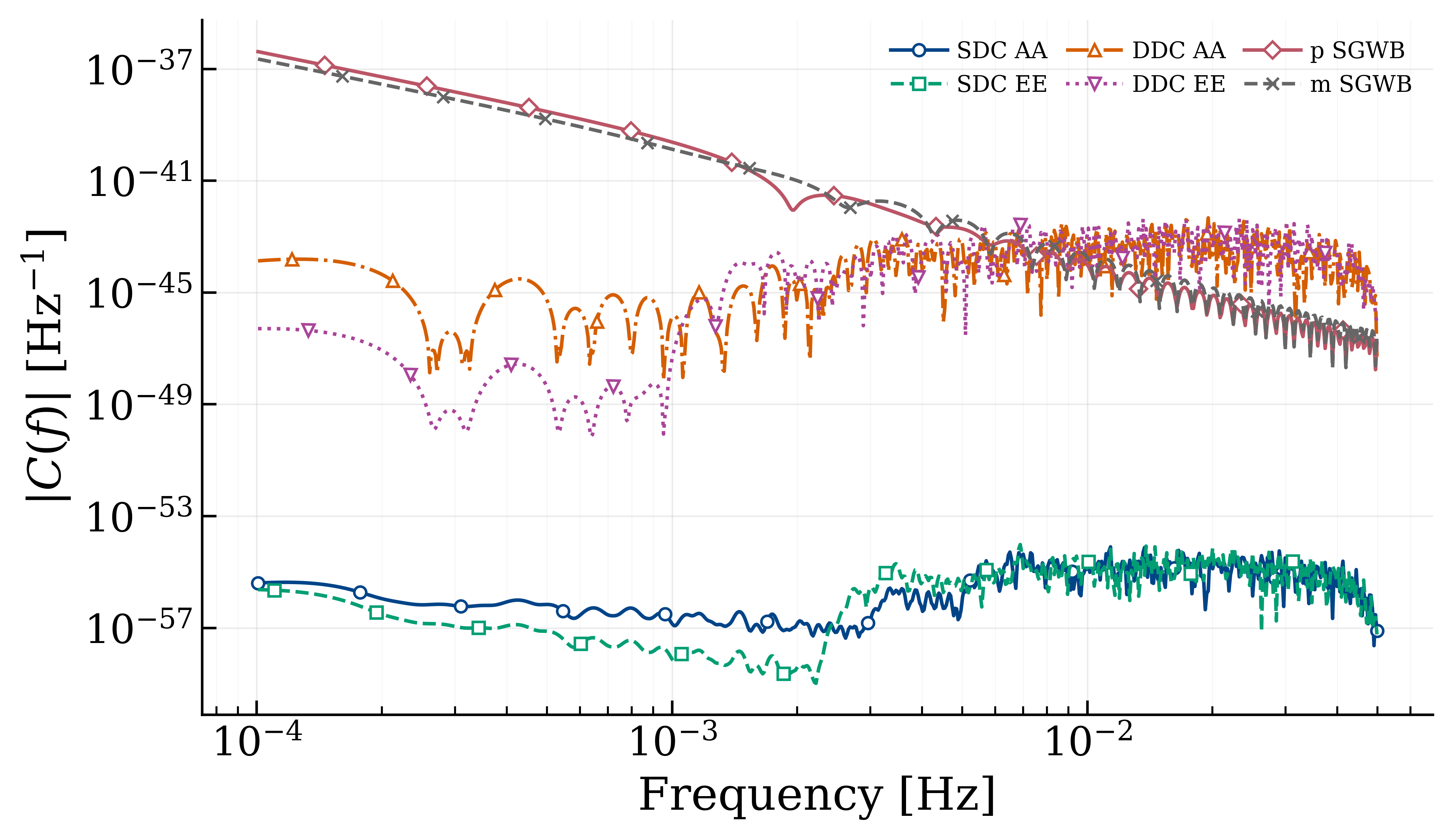}
\caption{Plasma cross spectra and SGWB cross correlation responses in the A/E channels of the LISA--Taiji network. Solid curves with circles and dashed curves with squares show the SDC AA and EE plasma cross spectra, respectively. Dash-dotted curves with upward-pointing triangles and dotted curves with downward-pointing triangles show the DDC AA and EE plasma cross spectra, respectively. Solid curves with diamonds and dashed curves with crosses show the SGWB A/E channel cross correlation responses for the LISA--Taijip and LISA--Taijim configurations, respectively. The plasma cross spectrum is substantially larger under DDC than under SDC, showing that larger scale solar wind structures enhance correlated propagation noise between the two detectors.}
\label{fig:crossspec}
\end{figure}

Equation~\eqref{eq:tdicross} first computes the full complex cross spectrum at the link level and then performs the TDI linear combination, retaining destructive and constructive interference among different links. Figure~\ref{fig:crossspec} shows the A/E cross spectra and SGWB response baselines for LISA--Taijip and LISA--Taijim. In both relative orbital configurations, the frequency structure of the cross spectrum is jointly determined by the finite arm response, TDI delay phases, and interdetector spatial correlation. The differences between A and E curves mainly arise from the different $\bm D_{I,d}$, while the differences between Taijip and Taijim mainly reflect the different geometric weights of the center baseline and link directions in $\Gamma_n$ and $\gamma_I$.

\section{Effect of Interdetector Plasma Noise on a Power Law SGWB}

We next evaluate the plasma noise amplitude, interdetector correlation strength, and Fisher parameter bias for a power law SGWB.

\subsection{Plasma noise amplitude and interdetector correlation strength}

The single detector noise level compares the plasma auto spectrum with the reference noise of that detector, whereas the impact on cross correlation estimation depends on the cross spectrum between the two detectors. To separate these two effects, we define the displacement or strain amplitude spectral density ratio
\begin{equation}
r_{I,d}(f)=
\sqrt{\frac{S_{I,d}^{\rm plasma}(f)}
{S_{I,d}^{\rm ref}(f)}} .
\label{eq:budgetratio}
\end{equation}
Table~\ref{tab:budget} shows that all listed frequencies have $r_{I,d}<1$, with the Taiji single link value at $3.3\,\mathrm{mHz}$ equal to $3.37\times10^{-2}$. The ratios in the A/E channels are affected by the TDI delay combination, but their order of magnitude remains below the reference noise. The plasma residual in a single Taiji detector is below the reference noise \cite{Xie2024}, but Eq.~\eqref{eq:budgetratio} contains only single detector auto spectrum information. The interdetector correlation must be determined from the network cross spectrum.

Table~\ref{tab:coh} reports the interdetector correlation strengths. At $1\,\mathrm{mHz}$, SDC yields link and TDI correlation strengths of $1.37\times10^{-11}$ and $6.16\times10^{-12}$, respectively. Under DDC, they reach $0.805$ and $0.757$. Equation~\eqref{eq:coherence} is normalized by the plasma auto spectra, so a high correlation strength indicates that the plasma components in the two detectors have a strong linear correlation. It does not mean that this component dominates over instrumental noise or the SGWB response. The single detector ratio, interdetector correlation strength, and parameter bias refer to different physical effects and are therefore discussed separately.

\begin{table*}[t]
\centering
\caption{Interdetector plasma correlation strengths for LISA--Taiji. The link correlation strength and TDI correlation strength are both evaluated at $1\,\mathrm{mHz}$.}
\label{tab:coh}
\begin{tabular}{lcccccc}
\hline\hline
Case & $L_\parallel$/km & $L_\perp$/km & $V_{\rm sw}$/km s$^{-1}$ & Correlation axis angle & Link correlation strength & TDI correlation strength\\
\midrule
SDC & $10^8$ & $10^7$ & 400 & $45^\circ$ & $1.37\times10^{-11}$ & $6.16\times10^{-12}$\\
DDC & $10^9$ & $10^8$ & 400 & $60^\circ$ & 0.805 & 0.757\\
\hline\hline
\end{tabular}
\end{table*}

\paragraph*{Equivalent SGWB amplitude shift.}
For an observing time $T$, the channel cross spectrum estimator for positive frequency data is
\begin{equation}
\widehat C_I(f)=\frac{2}{T}
\widetilde s_{I,\Taiji}^{\,*}(f)
\widetilde s_{I,\LISA}(f).
\label{eq:crossestimator}
\end{equation}
In the weakly correlated signal limit, its variance over a frequency bin $\Delta f$ is approximately
\begin{equation}
\operatorname{Var}\!\left[\widehat C_I(f)\right]
\simeq\frac{P_{I,\Taiji}(f)P_{I,\LISA}(f)}
{2T\Delta f}.
\label{eq:crossvariance}
\end{equation}
This yields the continuous frequency Gaussian likelihood
\begin{equation}
-2\ln\mathcal L=
2T\int\dd f\sum_{I=A,E}
\frac{\left|\widehat C_I(f)-C_I(f,\bm\theta)\right|^2}
{P_{I,\Taiji}(f)P_{I,\LISA}(f)}
+\mathrm{const}.
\label{eq:crosslikelihood}
\end{equation}
Under the weak cross signal and A/E orthogonal channel assumptions, the joint network frequency domain Gaussian likelihood reduces to Eq.~\eqref{eq:crosslikelihood} \cite{Wang2024PRD}. The subsequent parameter estimation and parameter bias calculation use the same frequency weights.

For a power law spectrum with fixed spectral index $Q(f)=(f/f_{\rm ref})^\alpha$, the model is $C_I^{\rm model}=\Omega_{\rm ref}G_I$, where $G_I=K\gamma_IQ$. We define the network inner product by
\begin{equation}
(a|b)=2T\int\dd f\sum_{I=A,E}
\frac{\operatorname{Re}[a_I^*(f)b_I(f)]}
{P_{I,\Taiji}(f)P_{I,\LISA}(f)} .
\label{eq:innerproduct}
\end{equation}
Maximizing Eq.~\eqref{eq:crosslikelihood} with respect to $\Omega_{\rm ref}$ yields the maximum likelihood estimator and its parameter uncertainty,
\begin{equation}
\widehat\Omega_{\rm ref}=
\frac{(G|\widehat C)}{(G|G)},\qquad
u_{\Omega_{\rm ref}}=(G|G)^{-1/2}.
\label{eq:omegaestimator}
\end{equation}
The corresponding network signal to noise ratio (SNR) is
\begin{equation}
\mathrm{SNR}^2=
\left(C^{\rm SGWB}|C^{\rm SGWB}\right)
:=\Omega_{\rm ref}^2(G|G),
\label{eq:networksnr}
\end{equation}
which defines the LISA--Taiji network SNR after summing over the A/E cross correlation responses \cite{Wang2023EPJC}.

\begin{figure}[t]
\centering
\includegraphics[width=\columnwidth]{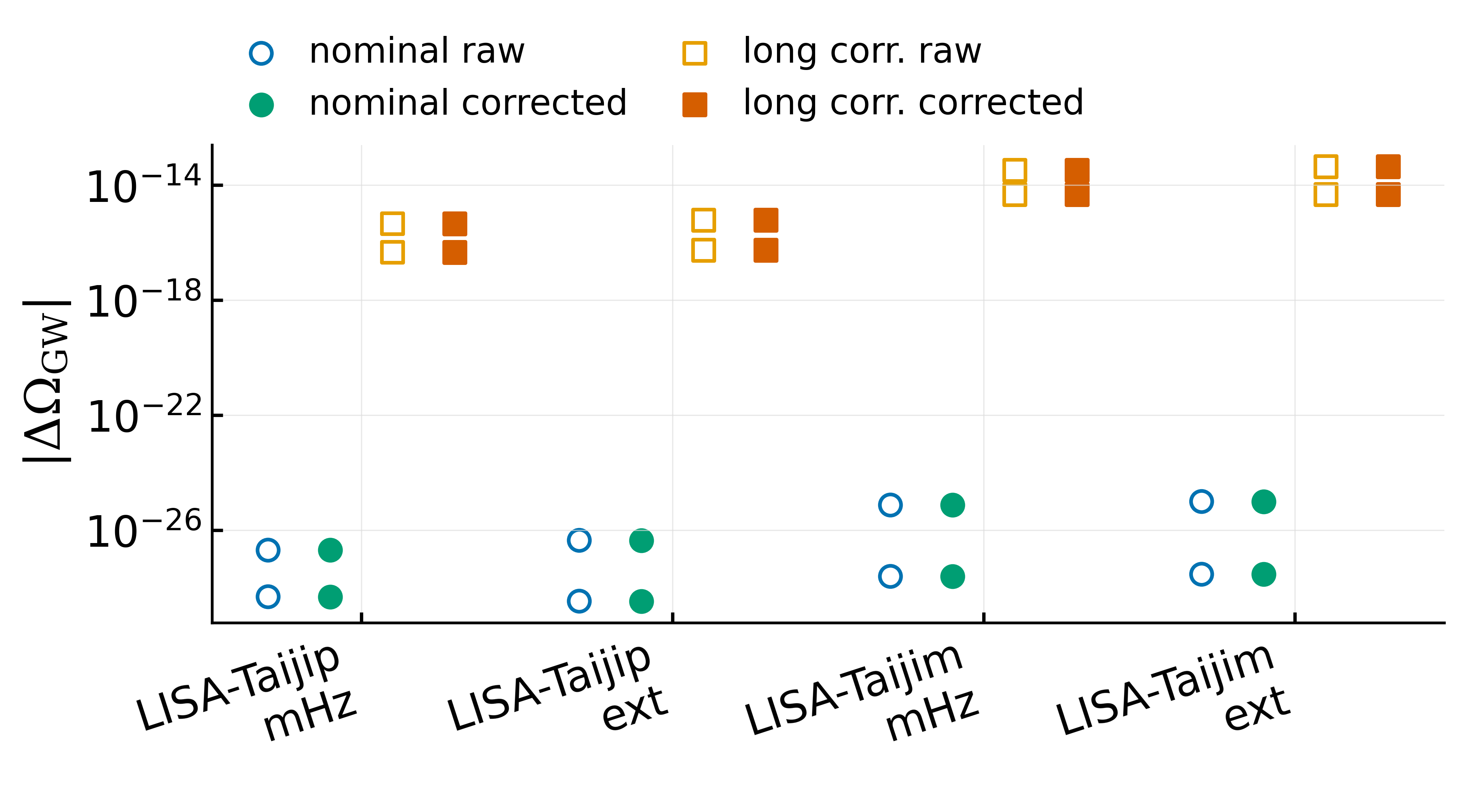}
\caption{Equivalent amplitude shifts $\Delta\Omega_{\rm GW}$ for the two network configurations and frequency bands. Open circles, filled circles, open squares, and filled squares denote the nominal raw, nominal corrected, long-correlation raw, and long-correlation corrected results, respectively. For each case, the lower and upper points show the magnitude of the signed estimate and the absolute upper bound, respectively.}
\label{fig:deltaomega}
\end{figure}

The quantity $\Delta\Omega_{\rm GW}$ denotes the equivalent amplitude shift obtained by projecting the plasma cross spectrum onto the SGWB amplitude under a fixed spectral shape. If the true cross spectrum also contains an unmodeled $C_I^{\rm plasma}$, the expectation value of $\widehat\Omega_{\rm ref}$ is shifted relative to the true amplitude. The corresponding $\Delta\Omega_{\rm GW}$ is
\begin{equation}
\Delta\Omega_{\rm GW}=
\frac{(G|C^{\rm plasma})}{(G|G)}.
\label{eq:deltaomega}
\end{equation}
Equation~\eqref{eq:deltaomega} uses the same response functions and noise weights as the SGWB amplitude estimator. Figure~\ref{fig:deltaomega} shows that SDC yields $\Delta\Omega_{\rm GW}<10^{-25}$ under the response weighted definition of Eq.~\eqref{eq:deltaomega}, whereas the DDC result for LISA--Taijim in the $10^{-4}$ to $10^{-2}\,\mathrm{Hz}$ band is $4.34\times10^{-14}$. Interdetector correlation produces a nonzero equivalent amplitude shift only when the plasma cross spectrum has a nonzero projection onto the SGWB response in frequency regions with large Fisher weight.

\subsection{Fisher parameter bias for a power law spectrum}

We take the parameters to be $\bm\theta=(\ln\Omega_{\rm ref},\alpha)$. In the weak signal approximation for cross correlations, the SGWB Fisher matrix and the bias vector generated by the unmodeled plasma cross spectrum are \cite{Fisher1922,CutlerFlanagan1994,Vallisneri2008,CutlerVallisneri2007}
\begin{align}
F_{ij}&=2T\int\dd f\sum_I
\frac{\operatorname{Re}\left[(\partial_i C_I^{\rm SGWB})^*
\partial_j C_I^{\rm SGWB}\right]}
{P_{I,\Taiji}P_{I,\LISA}},\\
b_j&=2T\int\dd f\sum_I
\frac{\operatorname{Re}\left[(\partial_j C_I^{\rm SGWB})^*
C_I^{\rm plasma}\right]}
{P_{I,\Taiji}P_{I,\LISA}},\\
\delta_{\rm pl}\theta_i&=(F^{-1})_{ij}b_j,\qquad
u_i=\sqrt{(F^{-1})_{ii}}.
\label{eq:fisherbias}
\end{align}
Here $F_{ij}$ is the Fisher matrix, $b_j$ is the perturbation vector generated by the plasma cross spectrum, $\delta_{\rm pl}\theta_i$ is the plasma induced parameter bias, and $u_i$ is the Fisher parameter uncertainty. The Fisher results are obtained from frequency integrals over the bands listed in the tables, rather than from a single displayed frequency. We define
\begin{equation}
\eta_{\theta_i}\equiv\frac{|\delta_{\rm pl}\theta_i|}{u_i}\times100\%
\label{eq:relativebias}
\end{equation}
as the plasma induced parameter bias in units of the corresponding Fisher parameter uncertainty. Below, we use $\eta_{\ln\Omega}$, $\eta_{\alpha}$, and $\eta_{\ln G\mu}$ for the corresponding parameters. The power law Fisher calculation uses the fiducial values $\Omega_{\rm ref}=10^{-12}$, $f_{\rm ref}=1\,\mathrm{mHz}$, and $\alpha=0$. The percentage for $\alpha$ is therefore not a relative error with respect to $\alpha$ itself, but the plasma induced bias in $\alpha$ divided by its Fisher parameter uncertainty. For the 1, 3, 5, and 10 yr cases, the dimensional bias is approximately fixed, whereas the normalization by the Fisher parameter uncertainty changes with observing time.

More generally, the Fisher matrix can be obtained from the curvature of the frequency domain Gaussian likelihood with respect to the parameters. For zero mean complex Gaussian data with covariance matrix $\bm{\mathcal C}(f,\bm\theta)$, the covariance Fisher matrix is \cite{Wang2024PRD,Vallisneri2008}
\begin{equation}
F_{ij}^{\rm G}=
T\int_0^\infty\dd f\,
\operatorname{Tr}\left[
\bm{\mathcal C}^{-1}
\frac{\partial\bm{\mathcal C}}{\partial\theta_i}
\bm{\mathcal C}^{-1}
\frac{\partial\bm{\mathcal C}}{\partial\theta_j}
\right].
\label{eq:fishergeneral}
\end{equation}
Equation~\eqref{eq:fisherbias} is the cross correlation form of Eq.~\eqref{eq:fishergeneral} in the weak cross signal limit, with one sided spectral normalization and the A/E orthogonal channel approximation. The corresponding network Fisher matrix is
\begin{equation}
F_{ij}^{\rm net}=F_{ij}^{(A)}+F_{ij}^{(E)} ,
\label{eq:fishernet}
\end{equation}
where each $F_{ij}^{(I)}$ is given by the integral for fixed channel $I$ in the first line of Eq.~\eqref{eq:fisherbias}. Here ``joint'' means summing the information in the two independent A/E channels within the same LISA--Taiji network. LISA--Taijip and LISA--Taijim are two alternative orbital configurations, and their results are listed separately; they are not added as simultaneous observing data. If a full mission orbit is introduced and divided into time segments $t_k$, one can further write $F_{ij}^{\rm all}=\sum_kF_{ij}^{\rm net}(t_k)$.

\begin{figure}[t]
\centering
\includegraphics[width=\columnwidth]{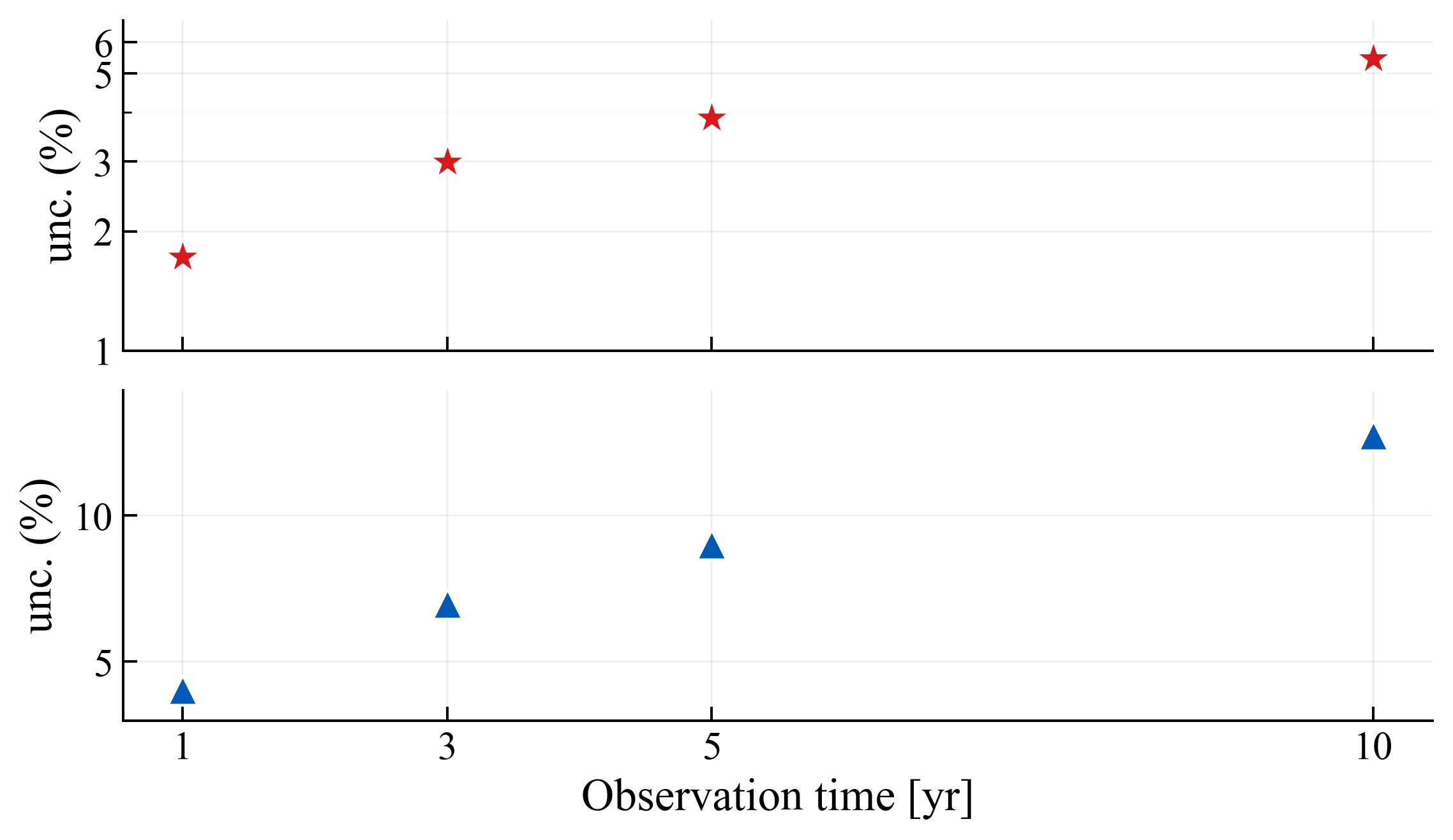}
\caption{Observation time dependence of power law parameter biases under DDC. The horizontal axis is the observing time, and the vertical axis is $\eta_\theta$. The upper and lower panels correspond to $\ln\Omega_{\rm ref}$ and $\alpha$, respectively.}
\label{fig:fisherbias}
\end{figure}

We use the power law spectral parameters
\begin{equation}
\bm\theta=(\theta_1,\theta_2)
:=(\ln\Omega_{\rm ref},\alpha).
\label{eq:params}
\end{equation}
From $C_I^{\rm SGWB}=K\gamma_I\Omega_{\rm ref}(f/f_{\rm ref})^\alpha$, we obtain
\begin{equation}
\begin{aligned}
\frac{\partial C_I^{\rm SGWB}}{\partial\ln\Omega_{\rm ref}}
&:=C_I^{\rm SGWB},\\
\frac{\partial C_I^{\rm SGWB}}{\partial\alpha}
&:=C_I^{\rm SGWB}\ln\left(\frac{f}{f_{\rm ref}}\right).
\end{aligned}
\label{eq:templatederivatives}
\end{equation}
If $\alpha$ is fixed and only the amplitude is estimated, the amplitude parameter uncertainty is
\begin{equation}
u_{\ln\Omega}^{\rm fixed}
:=\left(F_{\ln\Omega\ln\Omega}^{\rm net}\right)^{-1/2}.
\label{eq:singlefisher}
\end{equation}
When the amplitude and spectral index are jointly estimated, the full $2\times2$ Fisher matrix must be used,
\begin{equation}
\bm F=
\begin{pmatrix}
F_{\ln\Omega\ln\Omega}&F_{\ln\Omega\alpha}\\
F_{\ln\Omega\alpha}&F_{\alpha\alpha}
\end{pmatrix},\qquad
\bm\Sigma=\bm F^{-1}.
\label{eq:jointcovariance_short}
\end{equation}
The explicit inverse is
\begin{equation}
\bm\Sigma
:=\frac{1}{\det\bm F}
\begin{pmatrix}
F_{\alpha\alpha}&-F_{\ln\Omega\alpha}\\
-F_{\ln\Omega\alpha}&F_{\ln\Omega\ln\Omega}
\end{pmatrix}.
\label{eq:jointcovariance}
\end{equation}
The marginalized parameter uncertainties and parameter correlation coefficient are
\begin{equation}
u_{\ln\Omega}=\sqrt{\Sigma_{\ln\Omega\ln\Omega}},
\quad
u_\alpha=\sqrt{\Sigma_{\alpha\alpha}},
\quad
r_{\ln\Omega,\alpha}=
\frac{\Sigma_{\ln\Omega\alpha}}
{u_{\ln\Omega}u_\alpha}.
\label{eq:marginalerrors}
\end{equation}
Thus the results in Table~\ref{tab:fisher} correspond to joint constraints after marginalizing over the amplitude and spectral index. The off diagonal Fisher element and the correlation coefficient $r_{\ln\Omega,\alpha}$ describe the statistical correlation between $\ln\Omega_{\rm ref}$ and $\alpha$, so the marginalized parameter uncertainties must be obtained from the full covariance matrix. Since $F\propto T$ and the bias vector satisfies $b\propto T$, the dimensional parameter bias $\delta_{\rm pl}\theta=F^{-1}b$ is approximately independent of observing time. The Fisher parameter uncertainty satisfies $u_i\propto T^{-1/2}$, so the normalized bias $\eta_{\theta_i}$ increases as $T^{1/2}$. Figure~\ref{fig:fisherbias} and Table~\ref{tab:fisher} summarize the power law results.

\begin{table*}[t]
\centering
\caption{Fisher parameter bias results for the power law spectrum. The quantity $\eta_\theta$ is defined in Eq.~\eqref{eq:relativebias}. Values are obtained from integrals over the corresponding frequency bands and are listed for observing times of 1, 3, 5, and 10 yr.}
\label{tab:fisher}
\begingroup
\setlength{\tabcolsep}{10pt}
\begin{tabular}{llllrr}
\hline\hline
Case & Network & Band & Year & $\eta_{\ln\Omega}$ & $\eta_{\alpha}$\\
\midrule
\multirow{8}{*}{SDC} & \multirow{4}{*}{LISA--Taijip} & \multirow{4}{*}{$10^{-4}$ to $10^{-2}\,\mathrm{Hz}$} & 1 & $2.72\times10^{-14}\%$ & $6.85\times10^{-13}\%$\\
 &  &  & 3 & $4.71\times10^{-14}\%$ & $1.19\times10^{-12}\%$\\
 &  &  & 5 & $6.09\times10^{-14}\%$ & $1.53\times10^{-12}\%$\\
 &  &  & 10 & $8.61\times10^{-14}\%$ & $2.17\times10^{-12}\%$\\
\addlinespace[2pt]
 & \multirow{4}{*}{LISA--Taijim} & \multirow{4}{*}{$10^{-4}$ to $5\times10^{-2}\,\mathrm{Hz}$} & 1 & $1.12\times10^{-13}\%$ & $3.28\times10^{-13}\%$\\
 &  &  & 3 & $1.94\times10^{-13}\%$ & $5.68\times10^{-13}\%$\\
 &  &  & 5 & $2.50\times10^{-13}\%$ & $7.33\times10^{-13}\%$\\
 &  &  & 10 & $3.54\times10^{-13}\%$ & $1.04\times10^{-12}\%$\\
\midrule
\multirow{8}{*}{DDC} & \multirow{4}{*}{LISA--Taijim} & \multirow{4}{*}{$10^{-4}$ to $10^{-2}\,\mathrm{Hz}$} & 1 & $1.73\%$ & $4.03\%$\\
 &  &  & 3 & $2.99\%$ & $6.97\%$\\
 &  &  & 5 & $3.86\%$ & $9.00\%$\\
 &  &  & 10 & $5.46\%$ & $12.73\%$\\
\addlinespace[2pt]
 & \multirow{4}{*}{LISA--Taijim} & \multirow{4}{*}{$10^{-4}$ to $5\times10^{-2}\,\mathrm{Hz}$} & 1 & $1.69\%$ & $3.94\%$\\
 &  &  & 3 & $2.93\%$ & $6.83\%$\\
 &  &  & 5 & $3.78\%$ & $8.82\%$\\
 &  &  & 10 & $5.34\%$ & $12.47\%$\\
\hline\hline
\end{tabular}
\endgroup
\end{table*}

Under SDC, the Fisher parameter biases are below $10^{-11}\%$ of the corresponding parameter uncertainties. Under DDC, for LISA--Taijim in the $10^{-4}$ to $10^{-2}\,\mathrm{Hz}$ band, with 10 yr of observation and joint estimation of the amplitude and spectral index, the normalized bias for the spectral index reaches $\eta_\alpha=12.73\%$. Thus, if solar wind electron density structures can affect the links of both detectors simultaneously, interdetector plasma noise can induce an appreciable systematic bias in SGWB parameter estimation.

\section{Effect of Interdetector Plasma Noise on a Cosmic String SGWB}

We apply the same bias calculation to cosmic string M2/M3 spectra, network sensitivity, and the string tension parameter.

\subsection{Cosmic string spectra and network sensitivity}

In addition to the power law background, we use the cosmic string M2 and M3 spectra \cite{Wang2024PRD}. The M2 model includes contributions from loops formed in the radiation era, the radiation to matter transition, and the matter era, with $\alpha_{\rm loop}=0.1$, $\Gamma=50$, $\nu_r=1/2$, and $\nu_m=2/3$. The M3 model further includes contributions from small scale loops, with $\chi_r=0.2$, $\chi_m=0.295$, and a critical scale $\gamma_c=20(G\mu)^{1+2\chi}$. Both spectra are computed from the piecewise expressions in the literature, and the parameter $G\mu$ denotes the dimensionless string tension.

\begin{figure}[t]
\centering
\includegraphics[width=\columnwidth]{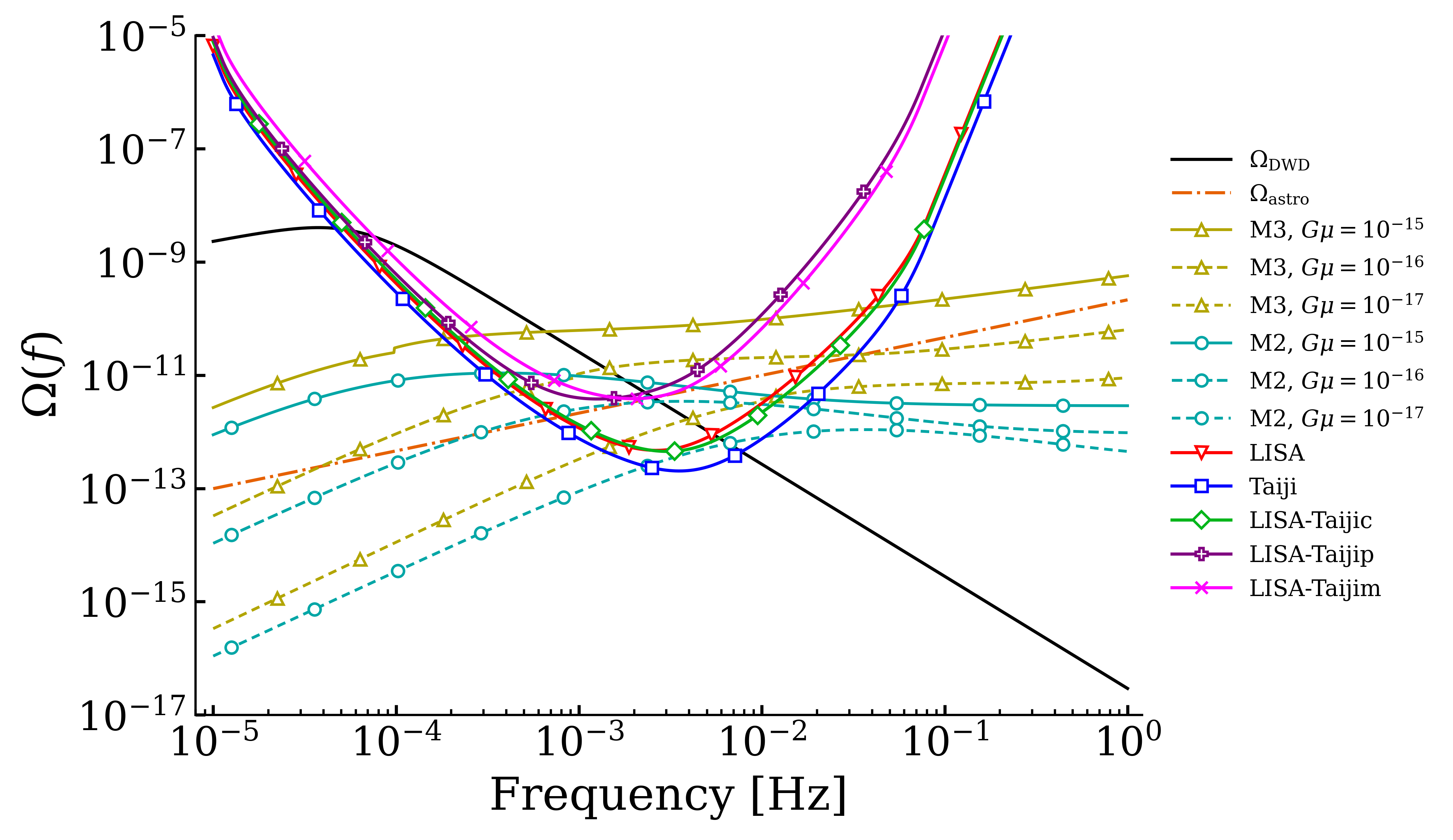}
\caption{Cosmic string M2/M3 backgrounds, Galactic double white dwarf foreground, extragalactic compact binary background, and 4 yr PLSs for different detector configurations. The M2/M3 curves use $G\mu=10^{-15},10^{-16},10^{-17}$, and the PLS threshold SNR is 10. Circles and upward-pointing triangles identify the M2 and M3 spectra, respectively. For each model, the three curves from top to bottom correspond to $G\mu=10^{-15},10^{-16},10^{-17}$. Downward-pointing triangles, squares, diamonds, plus signs, and crosses identify the PLSs for LISA, Taiji, LISA--Taijic, LISA--Taijip, and LISA--Taijim, respectively. The DWD and extragalactic compact binary backgrounds are shown by the solid and dash-dotted curves without markers, respectively.}
\label{fig:cosmicpls}
\end{figure}

To compare the signals, foregrounds, and network sensitivities in the same figure, we write the Galactic double white dwarf (DWD) foreground and the extragalactic compact binary background as
\begin{align}
\Omega_{\rm DWD}(f)&=
\frac{A_1(f/f_*)^{\alpha_1}}
{1+A_2(f/f_*)^{\alpha_2}},\\
\Omega_{\rm astro}(f)&=4.44\times10^{-12}
\left(\frac{f}{3\,\mathrm{mHz}}\right)^{2/3},
\label{eq:foregrounds}
\end{align}
where $A_1=7.44\times10^{-14}$, $A_2=2.96\times10^{-7}$, $\alpha_1=-1.98$, $\alpha_2=-2.6$, and $f_*=c/(2\pi L_{\rm LISA})$ \cite{Wang2024PRD}. For the power law family $\Omega_\kappa(f)=\Omega_\kappa(f_{\rm ref})(f/f_{\rm ref})^\kappa$ with spectral index $\kappa$, the normalization corresponding to the threshold SNR $\rho_{\rm th}$ is obtained from the network effective sensitivity $\Omega_n(f)$ as
\begin{equation}
\Omega_\kappa(f_{\rm ref})=
\frac{\rho_{\rm th}}
{\left[2T\int\dd f\,(f/f_{\rm ref})^{2\kappa}/\Omega_n^2(f)\right]^{1/2}}.
\label{eq:plsamp}
\end{equation}
The PLS is obtained by taking the envelope over $-8\leq\kappa\leq8$. Figure~\ref{fig:cosmicpls} uses $\rho_{\rm th}=10$ and $T=4\,\mathrm{yr}$, and shows LISA, Taiji, and the LISA--Taijip, LISA--Taijim, and LISA--Taijic configurations. LISA--Taijic is included only as a reference configuration for the network geometry. The plasma cross spectra are computed only for the p/m configurations with heliocentric separation baselines.

\begin{figure}[t]
\centering
\includegraphics[width=\columnwidth]{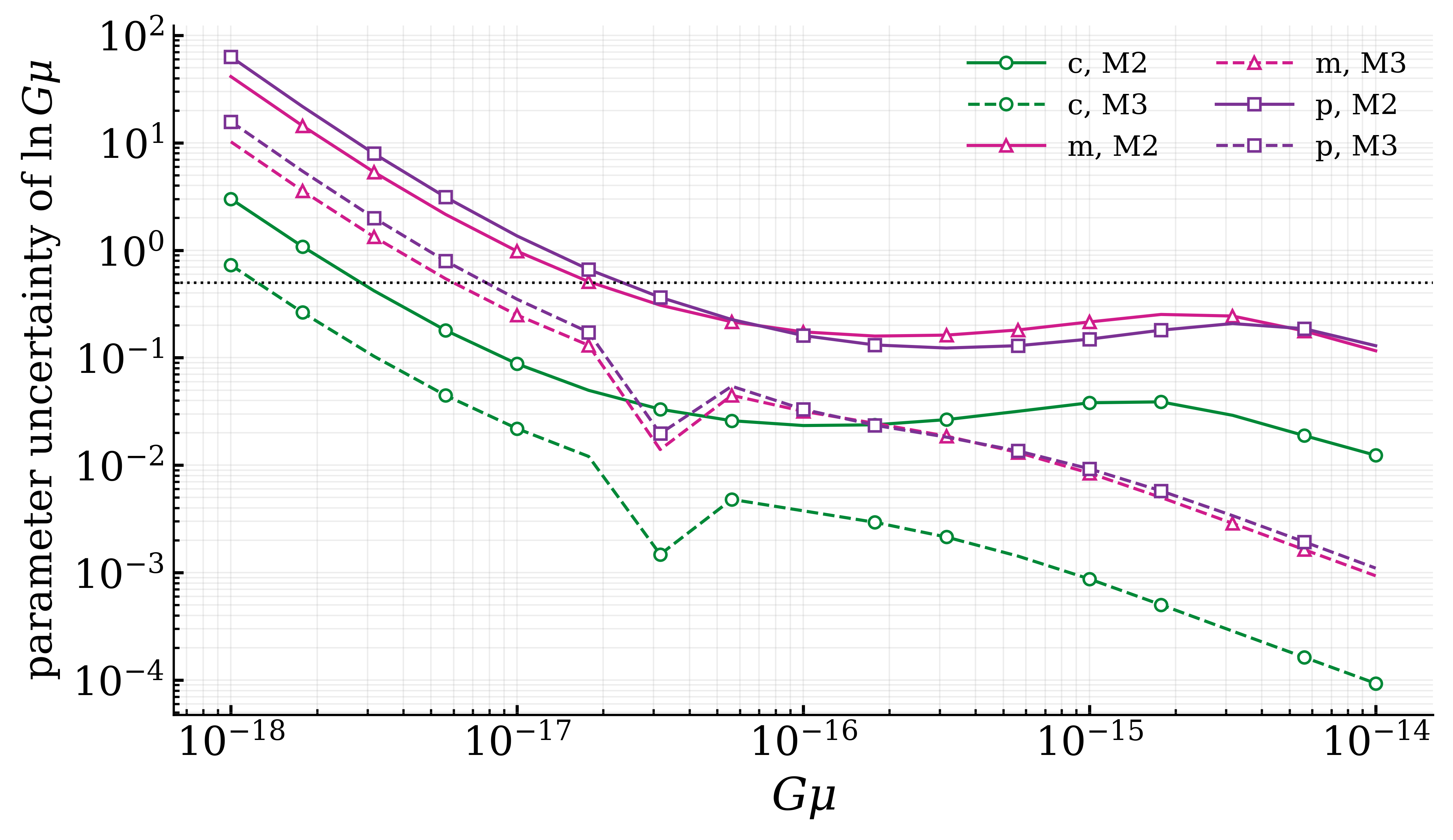}
\caption{String tension parameter uncertainties for the cosmic string M2/M3 spectra with 4 yr of observation. Solid and dashed curves denote the M2 and M3 models, respectively. Circles, squares, and upward-pointing triangles identify the LISA--Taijic, LISA--Taijip, and LISA--Taijim configurations, respectively.}
\label{fig:cosmicfisher}
\end{figure}

\subsection{Cosmic string parameter bias}

To isolate the role of network configuration from the power law spectral shape, we fix the M2 or M3 model and estimate only $\theta=\ln G\mu$. The spectral derivative in the cross correlation Fisher information of Eq.~\eqref{eq:fisherbias} is replaced by
\begin{equation}
\partial_{\ln G\mu}C_I^{\rm SGWB}
:=K(f)\gamma_I(f)\partial_{\ln G\mu}\Omega_{\rm cs}(f,G\mu),
\label{eq:gmuderivative}
\end{equation}
which yields $u_{\ln G\mu}=F_{\ln G\mu\ln G\mu}^{-1/2}$, where $u_{\ln G\mu}$ denotes the Fisher parameter uncertainty of $\ln G\mu$. Figure~\ref{fig:cosmicfisher} shows that LISA--Taijic, used as a reference network geometry, provides the strongest constraints across the full parameter range, whereas the p/m configurations have weaker string tension constraints because spatial separation suppresses the ORF.

\begin{figure}[t]
\centering
\includegraphics[width=\columnwidth]{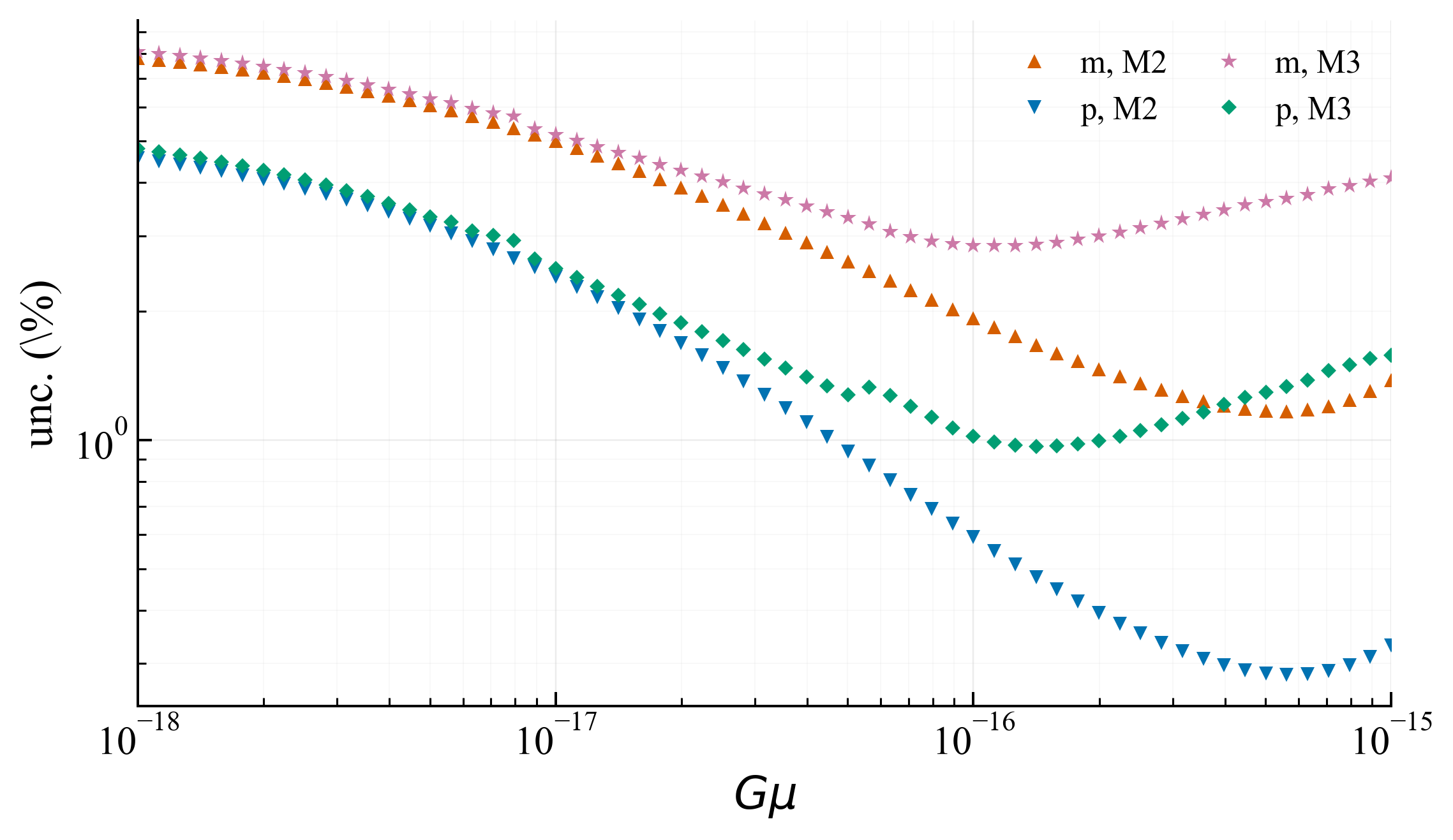}
\caption{Fisher parameter bias caused by the plasma cross spectrum for different string tensions. The horizontal axis is $G\mu\in[10^{-18},10^{-15}]$, and the vertical axis is $\eta_{\ln G\mu}$ under DDC.}
\label{fig:energybias}
\end{figure}

Figure~\ref{fig:energybias} shows the Fisher parameter bias caused by the plasma cross spectrum for different string tensions $G\mu$. The figure uses DDC and the $10^{-4}$ to $5\times10^{-2}\,\mathrm{Hz}$ band, scans $G\mu\in[10^{-18},10^{-15}]$, and summarizes the parameter bias percentages for 1, 3, 5, and 10 yr. In the cosmic string model, the SDC result is below $10^{-10}\%$, several orders of magnitude smaller than DDC. As $G\mu$ changes, the cosmic string spectral shape and Fisher weighting change simultaneously, so $\eta_{\ln G\mu}$ is not simply monotonic.

\begin{table}[t]
\centering
\caption{Fisher parameter bias results for M2/M3 cosmic string spectra in the LISA--Taijim configuration under DDC. The frequency band is $10^{-4}$ to $5\times10^{-2}\,\mathrm{Hz}$, and $\eta_{\ln G\mu}$ is defined by Eq.~\eqref{eq:relativebias}.}
\label{tab:energyfisher}
\begingroup
\setlength{\tabcolsep}{8pt}
\begin{tabular}{lcccc}
\hline\hline
Model & $G\mu$ & Year & $u_{\ln G\mu}$ & $\eta_{\ln G\mu}$\\
\midrule
\multirow{16}{*}{M2} & \multirow{4}{*}{$10^{-18}$} & 1 & 82.65 & $2.47\%$\\
 &  & 3 & 47.72 & $4.28\%$\\
 &  & 5 & 36.96 & $5.53\%$\\
 &  & 10 & 26.14 & $7.81\%$\\
\addlinespace[2pt]
 & \multirow{4}{*}{$10^{-17}$} & 1 & 1.958 & $3.75\%$\\
 &  & 3 & 1.131 & $6.49\%$\\
 &  & 5 & 0.876 & $8.37\%$\\
 &  & 10 & 0.619 & $11.84\%$\\
\addlinespace[2pt]
 & \multirow{4}{*}{$10^{-16}$} & 1 & 0.348 & $1.44\%$\\
 &  & 3 & 0.201 & $2.49\%$\\
 &  & 5 & 0.156 & $3.22\%$\\
 &  & 10 & 0.110 & $4.55\%$\\
\addlinespace[2pt]
 & \multirow{4}{*}{$10^{-15}$} & 1 & 0.430 & $1.03\%$\\
 &  & 3 & 0.248 & $1.79\%$\\
 &  & 5 & 0.192 & $2.31\%$\\
 &  & 10 & 0.136 & $3.27\%$\\
\midrule
\multirow{16}{*}{M3} & \multirow{4}{*}{$10^{-18}$} & 1 & 20.43 & $6.09\%$\\
 &  & 3 & 11.79 & $10.55\%$\\
 &  & 5 & 9.14 & $13.62\%$\\
 &  & 10 & 6.46 & $19.26\%$\\
\addlinespace[2pt]
 & \multirow{4}{*}{$10^{-17}$} & 1 & 0.498 & $3.83\%$\\
 &  & 3 & 0.288 & $6.63\%$\\
 &  & 5 & 0.223 & $8.56\%$\\
 &  & 10 & 0.157 & $12.11\%$\\
\addlinespace[2pt]
 & \multirow{4}{*}{$10^{-16}$} & 1 & 0.0634 & $2.13\%$\\
 &  & 3 & 0.0366 & $3.70\%$\\
 &  & 5 & 0.0284 & $4.77\%$\\
 &  & 10 & 0.0200 & $6.75\%$\\
\addlinespace[2pt]
 & \multirow{4}{*}{$10^{-15}$} & 1 & 0.0169 & $3.08\%$\\
 &  & 3 & 0.00975 & $5.33\%$\\
 &  & 5 & 0.00755 & $6.88\%$\\
 &  & 10 & 0.00534 & $9.72\%$\\
\hline\hline
\end{tabular}
\endgroup
\end{table}

Under DDC and in the $10^{-4}$ to $5\times10^{-2}\,\mathrm{Hz}$ band, the M3 spectrum in the LISA--Taijim configuration yields $\eta_{\ln G\mu}=19.26\%$ at $G\mu=10^{-18}$ and 10 yr of observation. Changing $G\mu$ changes the cosmic string spectral shape and therefore the Fisher projection of the plasma cross spectrum. For model spectra such as M2/M3, the parameter bias can reach the ten percent level as a fraction of the Fisher parameter uncertainty. The calculations in Figs.~\ref{fig:cosmicfisher} and \ref{fig:energybias}, and in Table~\ref{tab:energyfisher}, use the weak signal Fisher matrix with $\ln G\mu$ as the estimated parameter to compare configurations and evaluate correlated noise. This calculation differs from a full parameter Fisher analysis that jointly includes single detector auto spectra, cross spectra, and foreground parameters \cite{Wang2024PRD}.

Under SDC, the interdetector impact is suppressed by the scale separation among the millihertz frozen flow scale, the detector center separation, and finite arm averaging. Even if a single link accumulates a nonzero electron column density fluctuation, two detectors separated by about $10^8\,\mathrm{km}$ typically sample different plasma structures, and the frozen flow phase makes the cross spectrum between links vary rapidly with frequency. The TDI correlation strength in SDC is $6.16\times10^{-12}$, and the parameter bias for the power law spectrum under SDC is below $10^{-11}\%$ of the corresponding Fisher parameter uncertainty.

In DDC, where $L_\parallel$ and $L_\perp$ are larger, the two detectors can maintain a high correlation strength with the same large scale structure, and the TDI correlation strength reaches 0.757. Correlation strength alone does not determine the SGWB parameter bias; the ORF frequency structure, A/E response, frequency weighting, and instrumental noise jointly control the final result. In the power law test, the $\alpha$ parameter bias under DDC reaches $12.73\%$ of the corresponding Fisher parameter uncertainty.

The single detector noise level and interdetector plasma noise are different physical quantities. The former measures the contribution of the plasma residual to the noise level of Taiji or LISA itself, whereas the latter quantifies the parameter bias caused by the same environmental term in network cross correlation estimation. The single detector ratio $3.37\times10^{-2}$ in Table~\ref{tab:budget} and the network parameter bias below $10^{-11}\%$ under SDC can occur simultaneously. The single detector ratio alone cannot exclude correlated environmental noise. The same distinction also applies to other space detector networks and common environmental foregrounds \cite{Liang2022,Martinovic2021}.

The plasma cross spectrum produces a bias only when it has a nonzero Fisher weighted inner product with $\partial_{\ln\Omega}C^{\rm SGWB}$ or $\partial_\alpha C^{\rm SGWB}$. The derivative with respect to the amplitude parameter $\ln\Omega_{\rm ref}$ has the same spectral shape as the SGWB, whereas the derivative with respect to the spectral index parameter $\alpha$ has an additional factor $\ln(f/f_{\rm ref})$ and is therefore more sensitive to spectral shape changes on either side of the reference frequency. The parameter bias associated with $\alpha$ is larger in the DDC cases considered here, indicating that the residual cross spectrum has a stronger effect on the spectral slope parameter. The off diagonal elements in the joint Fisher matrix of amplitude and spectral index mix the two parameters, and the marginalized parameter uncertainties must be obtained from the full inverse matrix in Eq.~\eqref{eq:jointcovariance}.

The fully coherent arm approximation in Eq.~\eqref{eq:smetana_transfer} extends the Wind single point time series directly to the entire laser link and therefore predicts a larger displacement noise \cite{Smetana2020}. Finite arm averaging introduces an additional $f^{-1}$ filter in Eq.~\eqref{eq:jennrich_transfer}, reducing the single link estimate in the millihertz band \cite{Jennrich2021}. Applying the Lomb--Scargle periodogram to unevenly sampled Wind/SWE data and propagating the link noise to first generation TDI produces a Taiji plasma residual below the displacement noise requirement \cite{Xie2024}. Here we compute the nonzero block $\bm S_{\Taiji\LISA}^{\rm link}$ between the two sets of detector link spectral matrices, which characterizes interdetector plasma noise in SGWB searches.

These results show that the impact of solar wind plasma on LISA--Taiji SGWB searches cannot be determined from the single detector residual noise level alone. Even if the plasma auto spectrum is below the instrumental noise, cross correlation estimation acquires a parameter bias when the same solar wind density structure maintains a nonzero correlation between the two detectors' links and its cross spectrum projects onto the SGWB response in frequency regions with large Fisher weight. Under SDC, this projection is suppressed by spatial decoherence and the parameter bias is negligible. For the larger correlation lengths assumed in DDC, the plasma cross spectrum can induce a power law parameter bias at the ten percent level of the corresponding Fisher parameter uncertainty. Interdetector solar wind plasma noise is therefore a systematic term that must be tested separately in SGWB parameter estimation with the LISA--Taiji network; it cannot be ruled out solely because the single detector plasma noise power spectrum is below the instrumental noise power spectrum.

\section{Summary and Outlook}

Using Wind/SWE electron density data, we construct a spatial correlation model that includes finite arm propagation and Taylor frozen flow. We then propagate plasma noise to the TDI A/E channels and combine it with the LISA--Taiji ORF to compute the impact of interdetector solar wind plasma noise on SGWB cross correlation searches. Under SDC, both the interdetector correlation strength and the Fisher parameter bias remain small. If solar wind electron density structures can simultaneously affect LISA--Taiji links, the parameter bias for a power law SGWB can reach $12.73\%$ of the corresponding Fisher parameter uncertainty. The M2/M3 cosmic string results show that geometric differences between the p/m configurations and the LISA--Taijic reference configuration can shift the detection threshold for $G\mu$ by about one order of magnitude. Under DDC, the parameter bias for M2/M3 cosmic string spectra can reach $19.26\%$ of the corresponding Fisher parameter uncertainty.

Future work can use multipoint solar wind observations from Wind, Advanced Composition Explorer (ACE), Solar Orbiter, Parker Solar Probe, and other missions to constrain the dependence of $L_\parallel$ and $L_\perp$ on frequency, solar activity, and heliocentric distance. Such extensions could also incorporate full mission orbits and time dependent responses. Adding $L_\parallel$, $L_\perp$, $V_{\rm sw}$, and the normalization of the electron density spectrum to the parameter vector would enable joint Fisher or Bayesian modeling with both SGWB and environmental parameters. This framework could be extended to dual detector or triple detector networks involving TianQin. It can also be applied to first order phase transition backgrounds, scalar induced backgrounds, and other early Universe models with peaked spectral shapes, where the relative position between the peak frequency and the LISA--Taiji sensitive band affects the Fisher projection of the plasma cross spectrum. Future work may also include the Galactic white dwarf foreground and compact binary background to test how interdetector solar wind plasma noise affects SGWB component separation in the LISA--Taiji network \cite{Lamberts2019,Perigois2021,Boileau2022,Auclair2020,Caprini2016,Bartolo2016}.

\begin{acknowledgments}
This work was supported by the National Key Research and Development Program of China (Grant No.~2023YFC2206702), the Fundamental Research Funds for the Central Universities Project (Grant No.~2024IAIS-ZD009), the National Natural Science Foundation of China (Grant Nos.~12575072, 12547101 and 125B2102), and the Natural Science Foundation of Chongqing (Grant No.~CSTB2023NSCQ-MSX0103).
\end{acknowledgments}

\nocite{*}
\clearpage
\bibliography{wenxian}

\end{document}